\documentclass{article}
\usepackage[utf8]{inputenc}
\usepackage{amsmath}
\usepackage{amsfonts}
\usepackage{amssymb}
\usepackage{algorithm}
\usepackage{algpseudocode}
\usepackage{multirow}
\usepackage{float}
\usepackage{array}
\usepackage[left=0.75in,right=0.75in,top=0.75in,bottom=0.75in]{geometry}
\usepackage{comment}
\usepackage{graphicx}
\usepackage{url}
\usepackage{natbib}
\usepackage{color}
\usepackage{pgfplots}
\usepackage{caption}
\usepackage{subcaption}
\usepackage{pdflscape}
\usepackage{booktabs} 
\usepackage{enumerate}
\usepackage{authblk}
\usepackage{blkarray}
\usepackage{bm}
\usepackage{marvosym} 
\usepackage{soul}
\usepackage{textcomp, xspace}
\usepackage{mathtools}
\usepackage{tfrupee}
\usepgfplotslibrary{dateplot}
\usepackage[colorlinks]{hyperref}
\hypersetup{bookmarks=true, citecolor=blue, urlcolor=blue, linkcolor=blue}

\definecolor{greenxllite}{RGB}{196,215,155}
\definecolor{bluexllite}{RGB}{149,179,215}
\definecolor{redxllite}{RGB}{218,150,148}
\definecolor{greenxl}{RGB}{155, 187, 89}
\definecolor{bluexl}{RGB}{79,129,189}
\definecolor{redxl}{RGB}{192, 80, 77}
\definecolor{OliveGreen}{RGB}{70,136,52}

\usetikzlibrary{patterns}
\allowdisplaybreaks

\newcolumntype{x}[1]{>{\centering\arraybackslash\hspace{0pt}}p{#1}}

\newcommand{\Keywords}[1]{\par\noindent
{\small{\em \textbf{Keywords}\/}: #1}}

\title{An Agent-Based Fleet Management Model for \\ First- and Last-Mile Services}

\author[1] {Saumya Bhatnagar}
\author[1,2] {Tarun Rambha} 
\author[3] {Gitakrishnan Ramadurai} 

\affil[1]{\small Centre for infrastructure, Sustainable Transportation and Urban Planning (CiSTUP), Indian Institute of Science (IISc), Bengaluru, India}
\affil[2]{\small Civil Engineering, Indian Institute of Science (IISc), Bengaluru, India}
\affil[3]{\small Department of Civil Engineering, Indian Institute of Technology Madras (IITM), Chennai, India}

\date{ }

\begin{document}
\maketitle
\vspace{-5mm}
\begin{abstract}
With the growth of cars and car-sharing applications, commuters in many cities, particularly developing countries, are shifting away from public transport. These shifts have affected two key stakeholders: transit operators and first- and last-mile (FLM) services. Although most cities continue to invest heavily in bus and metro projects to make public transit attractive, ridership in these systems has often failed to reach targeted levels. FLM service providers also experience lower demand and revenues in the wake of shifts to other means of transport. Effective FLM options are required to prevent this phenomenon and make public transport attractive for commuters. One possible solution is to forge partnerships between public transport and FLM providers that offer competitive joint mobility options. Such solutions require prudent allocation of supply and optimised strategies for FLM operations and ride-sharing. To this end, we build an agent- and event-based simulation model which captures interactions between passengers and FLM services using statecharts, vehicle routing models, and other trip matching rules. An optimisation model for allocating FLM vehicles at different transit stations is proposed to reduce unserved requests. Using real-world metro transit demand data from Bengaluru, India, the effectiveness of our approach in improving FLM connectivity and quantifying the benefits of sharing trips is demonstrated.
\end{abstract}

%\begin{keyword}
%transportation \sep first- and last-mile services \sep agent-based modelling \sep
%multi-modal integration \sep 
%shared mobility \sep 
%\end{keyword}


%\end{frontmatter}
%\linenumbers

\section{Introduction}
Despite the growth of mass rapid transit systems, a large segment of commuters continues to travel by personal vehicles in many metropolitan cities. According to a study by World Resources Institute and Toyota Mobility Foundation---STation Access and Mobility Program (STAMP)---one of the significant challenges commuters face while using transit, such as metro trains, is the lack of first- and last-mile (FLM) connectivity \citep{STAMP:2017}. Moreover, due to ride-hailing alternatives such as Uber, DiDi, and Ola, there has been a significant mode shift from public transportation to personalised door-to-door services. These shifts have not only resulted in an increase in the usage of motorised vehicles but have also reduced revenue streams for FLM services and have led to the under-utilisation of public transport. Therefore, to improve transit ridership and make it more accessible to users, seamless integration with efficient FLM services is paramount.  

While extensive work has been carried out to evaluate on-demand autonomous vehicle (AV) services as an alternative to public transit \citep{winter2018performance, berrada2021economic}, work on FLM connectivity between public transport and on-demand services is relatively limited. Network- and agent-based simulation studies by \cite{mueller2011simulation}, \cite{martinez2015agent}, and \cite{fagnant2018dynamic} propose models for personal rapid transit systems and ride-sharing but do not offer insights on improving the efficiency of integrated public transport and FLM mobility services. Our current work aims to bridge this gap.

This research aims to build a multimodal simulation system that addresses critical planning and operational questions related to FLM connectivity. We demonstrate the benefits of this simulator using a detailed case study of metro networks and on-demand taxi-like FLM services such as auto-rickshaws that are popular in many developing countries, including India. The city of Bengaluru, for instance, had nearly 220,000 registered auto-rickshaws as of May 2020 \citep{Karnataka_transport}. Many private companies such as Cab4You, Tummoc, and Rapido offer app-based on-demand services in Bengaluru, which can be used for FLM connectivity \citep{Cab4u, Tummoc_Rapido}. Our proposed approach, however, is not specific to the type of FLM vehicles and can also be used for cars and vans. A commercial software \textit{AnyLogic} is employed with an additional customised codebase written for the fleet management problem and is tested using real-world data. The proposed simulation platform has the following features:
\begin{itemize}
\itemsep 0pt
    \item We fuse entry and exit hourly-ridership data from stations of Bengaluru metro, arrival schedules of trains, OpenStreetMaps (OSM), and population characteristics to approximate potential origins and destinations of travellers.  
    \item The simulator uses a combination of agent- and event-based modelling. Event-based models replicate train and passenger arrivals. On the other hand, agent-based models help create passenger and FLM agents and simulate their behaviour using \textit{statecharts}. Statecharts codify rules that allow the system and the agents to transition from one state to another.
    \item The simulator is designed to provide various performance metrics such as lost demand and utilisation rates at fine spatio-temporal scales that can be used to find planning and operational strategies that improve the system's efficiency. 
\end{itemize} 

Our simulation platform addresses FLM connectivity problems by answering two main questions. 
\begin{itemize}
    \item \textit{Planning Problem---How can the FLM fleet be optimally allocated across the network to improve system performance, such as lost demand?}

At the planning stage, we suppose that a fixed number of FLM vehicles, pre-assigned to a particular station, serve only those customers who board and deboard at that station and travel from/to nearby locations. Imposing such operational restrictions on FLM services avoids supply imbalances created due to the accumulation of FLM vehicles at metro stations with higher footfall and limits the empty miles travelled. Figuring out how many such vehicles to assign at each station is of interest to service providers and transit operators to minimise lost demand and consequently improve ridership and revenue. We estimate the lost demand based on the number of passengers who have to wait beyond a certain threshold due to the unavailability of an FLM vehicle. Additionally, the solution to this stage also helps to set aside parking spaces. To solve this problem, one could use optimisation engines that are integrated with simulation software (e.g., OptQuest in AnyLogic). However, such off-the-shelf tools would take several hours to evaluate the performance of a single feasible solution, especially when ride-sharing is allowed. Instead, we exploit the fact that the problem can be decomposed by stations and propose a simple but effective knapsack-like formulation with an approximate piecewise linear objective. The computational tractability of the optimisation formulation was improved using an iterative scheme which adaptively refines the feasible region. In addition, we also estimate the FLM service provider's revenue under different trip- and distance-based pricing schemes and comment on the parking needs at different stations. 

\item \textit{Operational Problem---How do operational strategies pertaining to sharing FLM rides impact key performance metrics?}

Sharing rides, in theory, can improve vehicle utilisation and the number of requests served. To quantify these benefits, we also simulate scenarios where FLM trips are shared and compare the lost demand with the no-sharing case. For the last-mile setting, FLM vehicles are assumed to pick up passengers from metro stations and drop them off at their respective destinations. Since passenger arrivals occur in batches, we formulate this scenario as a capacitated vehicle routing problem (CVRP) where each FLM vehicle is assumed to have a capacity of three. On the other hand, sharing of first-mile trips is more dynamic and is assumed to occur in a First-In-First-Out (FIFO) manner where a new request is inserted only after the pickup of the current request and if a detour time threshold is not exceeded. 

\end{itemize}

The rest of the paper is structured as follows. Section \ref{sec:litreview} summarises related literature on multimodal transport and FLM connectivity. In Section \ref{sec:methodology}, we describe the proposed architecture. An agent-based approach is presented with different modelling constructs for passenger and FLM agents and the optimisation model for resource allocation. We carry out a thorough evaluation of various performance indicators to measure the effectiveness of our approach in Section \ref{sec:experiments}. Section \ref{sec:conc} concludes this paper and provides directions for building on this research.

\section{Literature Review}
\label{sec:litreview}
Shared transport services can be broadly categorised into Fixed Transit (FT) and Demand Responsive Transport (DRT). FT uses high-capacity public transport services that follow fixed schedules and routes. They are, therefore, known to be cost-effective. However, in many cases, they are less convenient from a passenger's standpoint as the FLM legs of the main-haul trip need to be undertaken by other modes, such as bicycles, walking, or other personal vehicles. DRT, on the other hand, provides door-to-door services to individual passengers and has a flexible schedule and route.    

The third kind of hybrid transport system that integrates FT with DRT has the potential to offer advantages of both services \citep{lucken2019three}. In such systems, the main part of the journey is carried out via a bus or a high-speed train, while the FLM part of the trip is done using demand-responsive vehicles. An assessment framework to evaluate the increase in accessibility when DRT is used in the first/last leg of an FT journey was provided by \cite{alonso2018potential}. For feeder transit services, \cite{li2010feeder} proposed simulation and analytical models that planners can use to decide when FT and DRT policies should be used. The policy choice is based on service quality that includes passenger waiting, walking, and ride time. \cite{edwards2013comparing} developed comparison techniques to determine scenarios in which demand-responsive feeder services can be utilised effectively to lower operating costs and increase customer satisfaction. Analytical methods have also been proposed by \cite{chang1991integration}, \cite{aldaihani2004network}, and \cite{goswami2021} to integrate fixed route services with flexible demand-responsive services. 

Several studies have also been conducted on multimodal transport systems. \cite{shaheen2016mobility} provided a detailed analysis of different shared mobility modes, including on-demand rides, and their potential impact on vehicle miles travelled. \cite{kanuri2019leveraging} demonstrated significant time savings and modal shift from personal vehicles by integrating public transport with feeder services that improve last-mile connectivity. A study comparing the competing and complementing nature of taxi trips and public transit was done by \cite{wang2019new}. Their findings suggest that taxis compete with transit in areas with good public transportation, whereas a complementary effect can be observed in peripheral areas with low population density.
 
A few studies have also explored how emerging mobility modes such as electric and AVs can enhance multimodal travel. \cite{shen2018integrating} proposed a system that integrates AVs and public transport based on the demand characteristics of Singapore. They use an agent-based model for first-mile service simulation and evaluate the effectiveness of their approach by varying fleet size and sharing preferences. Key performance measures in their work include out-of-vehicle travel time, passenger car unit kilometres, operating cost, and revenue. Their findings indicate that such a system could be beneficial for low-demand bus routes, and as the demand increases, the system performance could worsen. Another related research was conducted by \cite{scheltes2017exploring}, where they designed a system called automated last-mile transport, consisting of electric vehicles that cater to last-mile connectivity of train trips. Their results show a reduction in travel time and waiting time with strategies such as intermediate charging of electric vehicles, short-term pre-booking, and relocating empty vehicles. \cite{LEFFLER2021103401} described a simulation framework for autonomous vehicles that integrates DRT with FT. They proposed a heuristic to rank trip plans based on cumulative passenger waiting times and the number of requests assigned. The operational cost and the level of service for passengers were also evaluated. Another study by \cite{STIGLIC201812} showed a reduction in total vehicle kilometres travelled when ride-sharing services are used in conjunction with mass transit.

A related problem to FLM routing is that of taxi dispatching and management of on-demand vehicle fleets. For shared taxi services, \cite{martinez2015agent} proposed an agent-based simulation model where a central dispatching algorithm matches incoming requests with available taxis to minimise extra travel time and waiting time, and maximise the revenue of taxis per kilometre travelled. \cite{mueller2011simulation} analysed the impact of different demand profiles and operational strategies for improving energy consumption, fleet utilisation, and system costs using a discrete-event personal rapid transport simulation model. \cite{fagnant2018dynamic} extended agent- and network-based simulations to include dynamic ride-sharing in shared AVs and studied its role in reducing average service time and travel costs. \cite{hartleb2021vehicle} proposed efficient heuristics for a fleet-sizing problem that serves demand estimated from a macroscopic model.

\cite{GRAHN2021103430} proposed a heuristic-based approach for demand matching and vehicle routing to cater to on-demand FLM requests. They evaluated their model using different operational policies and performance measures such as travel time and reliability. \cite{KUMAR2021102891} proposed a rolling horizon-based optimisation method to solve the ride-sharing matching problem in multimodal network settings. They maximised the number of matches between riders and drivers and the savings in the total vehicle hours. An FLM connectivity case study for Austin light rail with shared AVs was proposed by \cite{huang2021use}. A simulation toolkit was developed along with an implementation of a nested logit model for different mode choices such as walking, private cars, and walk-and-ride. Other related work includes simulation-based studies by \cite{costa2021simulation} and \cite{lau2021shared} that focus on improving FLM connectivity of public transit systems by integrating them with DRT. A simulation-based decision support tool was also proposed by \cite{segui2019simulating}, which combines demand generation models and a fleet simulator. The impact of various operational factors, such as vehicle fleet size, occupancy, and service levels, including passenger wait time and detour time, were studied.

Many of these agent-based studies do not consider the complete integration of public transport, particularly the schedules, for improving FLM connectivity. Moreover, they evaluate various scenarios to answer what-if questions but do not use optimisation to improve performance measures. In this study, we use real-world metro transit and census data of Bengaluru city and capture time-of-day variations in schedules and ridership. Our model also incorporates stochasticity in the number of passengers arriving at every station (based on historical data), passenger origins, and destinations. We also formulate an optimisation model to determine the optimal allocation of vehicles at different metro stations such that the number of unserved passenger trip requests is minimised under different ride-sharing scenarios.

\section{Methodology}
\label{sec:methodology}
\subsection{Dataset and Network Structure}
\begin{figure}[H]
    \centering
    \includegraphics[scale=1.6]{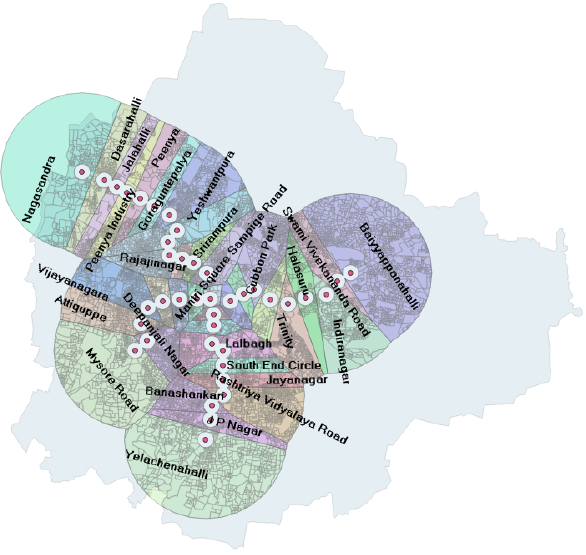}
    \caption{Metro transit network and Voronoi regions. The EBs and the Bengaluru city boundary are shown in the background.}
    \label{fig:transit_network}
\end{figure}

The data used in this study is from the Bengaluru city metro transit network that consists of 40 stations on the East-West corridor (Purple Line) and the North-South corridor (Green Line). The East-West corridor extends from Baiyappanahalli in the East to Mysore Road terminal in the West, while the North-South corridor extends from Nagasandra in the North to Yelachenahalli in the South as shown in Figure \ref{fig:transit_network}. The data includes train start times at these four reference stations and headway that varies by the time of the day. Using this information and the time taken for a train to travel from one station to the next, we estimate the train arrival times at every station. During peak hours, the frequency of metro trains is in the range of 3--6 minutes, while the off-peak frequency is 15--30 minutes. The average ridership per day is approximately 726,000 (pre-COVID).   

The transit demand data consists of monthly count of passengers entering and exiting these 40 stations for different hours of the day, from 5 AM to 11 PM. We first compute the number of passengers alighting at a station from each train by dividing the demand data by the number of days in the month. We assume that 10\% of this demand equals the mean value of the last-mile passenger count. The actual number of last-mile passengers arriving at different stations is simulated according to a Poisson distribution. For the first-mile scenario, we use 10\% of the hourly count of passengers entering the metro stations (as obtained from historical data) and generate their departure times within each hour using a uniform distribution. For passenger origins and destinations, we use census data of Bengaluru consisting of enumeration blocks (EBs) and their population. The centroids of these EBs serve as proxies for locations of first- and last-mile trips, respectively. The probability of choosing a location (centroid) is assumed to be proportional to the population of the EB to which it belongs. Currently, there are no centralised FLM providers, and hence we did not have data on the exact percentage of riders who might be interested in such services. We also could not simulate different trip types (e.g., work-based, recreational), which produce different patterns for FLM demand and their origin/destination locations. In the future, this kind of data could be sourced from surveys to get more realistic estimates. 

Only those EBs in Voronoi regions within a radius of 500 m to 5 km from the metro stations are considered. The Voronoi polygon of a particular station contains points closest to it (measured using the haversine distance) compared to any other station. If the roadway network is sparse, the shortest path lengths along the network could be used instead of haversine distances while constructing the Voronoi regions. For distances less than 500 m, we assume that passengers prefer walking, and for locations farther than 5 km, passengers are assumed to use other direct modes of travel. FLM vehicles assigned to a metro station are only assumed to serve locations in the corresponding Voronoi polygon. 

We extract the metro network, i.e., the two transit lines and 40 metro stations, using the Overpass API of OpenStreetMap, and convert it into GeoJSON and shapefiles. Figure \ref{fig:transit_network} shows a QGIS snapshot of the Bengaluru metro transit network and the Voronoi regions along the Purple and Green Line. The enumeration blocks are shown in grey colour. The outer boundary on the map indicates the city planning limits. The Nagasandra station, located at the end of the North-South corridor, can serve more demand from the city's suburbs, but the EB data was available only within the city limits.

\begin{figure}[ht]
    \centering
    \includegraphics[scale=0.3]{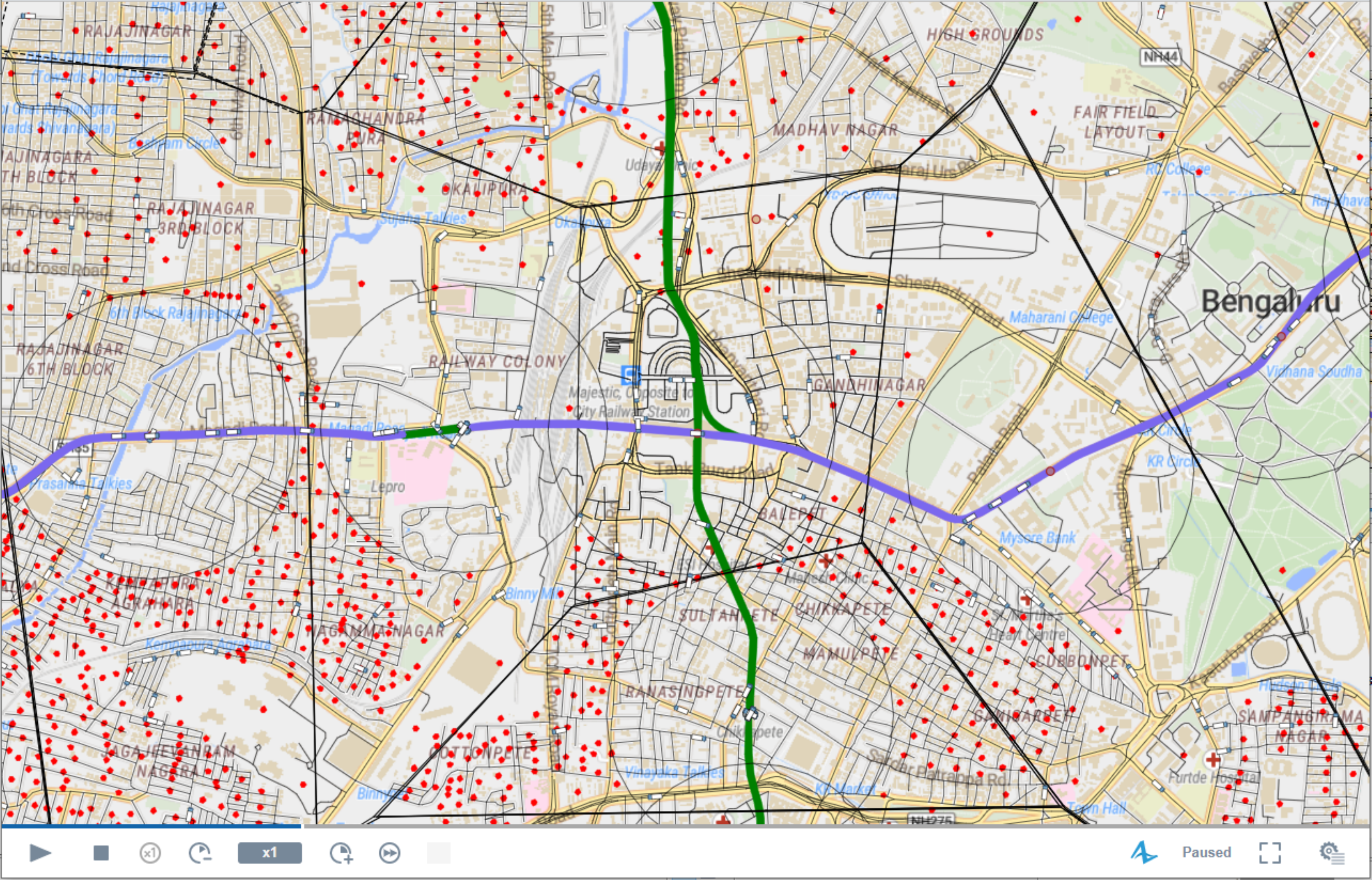}
    \caption{AnyLogic simulator showing the OSM network and Voronoi regions. Purple and green lines are the metro routes, and red dots represent centroids of EBs, which are potential origins/destinations. The circles indicate regions which are within 500 m.}
    \label{fig:simulator}
\end{figure}

Figure \ref{fig:simulator} shows a screenshot of the AnyLogic simulator. The FLM vehicles are routed using the shortest paths from the OSM server. The travel times on the road network are assumed constant based on the average speed of vehicles in Bengaluru and the length of the links.     

We currently do not assume time-of-day variations in network travel times. The waiting time calculation begins when a passenger makes an FLM request until a vehicle is available for boarding. Walking time to the parking lot is ignored. However, extending the simulator to capture these features using more realistic data is straightforward.

\subsection{Modelling Framework}
This section describes the model components of the proposed simulator. We use a combination of agent-based and discrete-event modelling paradigms. Agent-based modelling involves multiple decentralised entities (called \textit{agents}) who interact with each other within a simulation environment \citep{crooks2012introduction}. These agents represent real-world entities and can either be fixed in position (such as metro stations) or exhibit movement (such as people and vehicles). Agent behaviour is modelled using statecharts where each state represents a situation during the life cycle of an agent \citep{DBLP:reference/ap/GreenlawL02}. When an agent is in a specific state, it can perform a certain action(s) or wait for an event or timeout. The properties of agents can be specified using \textit{parameters} and \textit{variables}. Any action agents take within a state can be modelled using \textit{functions}. A state change occurs at the onset of a certain event or when a certain condition holds and is modelled using \textit{transitions}. Transitions can be triggered by a timeout, rate, message, agent arrival, or when a certain user-defined condition is true. Communication between agents is modelled using \textit{messages}. The message object and destination agent to whom the message is to be sent must be provided while designing the state chart. As soon as the destination agent receives the message, a set of necessary actions can also be specified. Additional details specific to AnyLogic are provided in Appendix \ref{appendix:specs}.

Figure \ref{fig:process_overview} shows an overview of the simulation architecture, including the different data sources used and the flow of execution for both first- and last-mile scenarios. Table \ref{tab:model_agents} gives a brief summary of agents used in the proposed model. All agents reside in \texttt{Main}, which can be viewed as the top-level agent. For the last-mile scenario, we define the timestamps at which trains arrive at the reference stations. At each of these timestamps, an event \texttt{eventTrainArrival} is triggered, which creates an instance of the agent type \texttt{MetroTrain}. This event can occur at an absolute model time or a specific date and time, as in our case. Next, an event \texttt{eventPassengerArrival} gets triggered at each subsequent metro station at the specified arrival time. For the first-mile case, the \texttt{eventPassengerArrival} is triggered randomly on an hourly basis. 

\begin{figure}[H]
\centering
\includegraphics[scale=0.52]{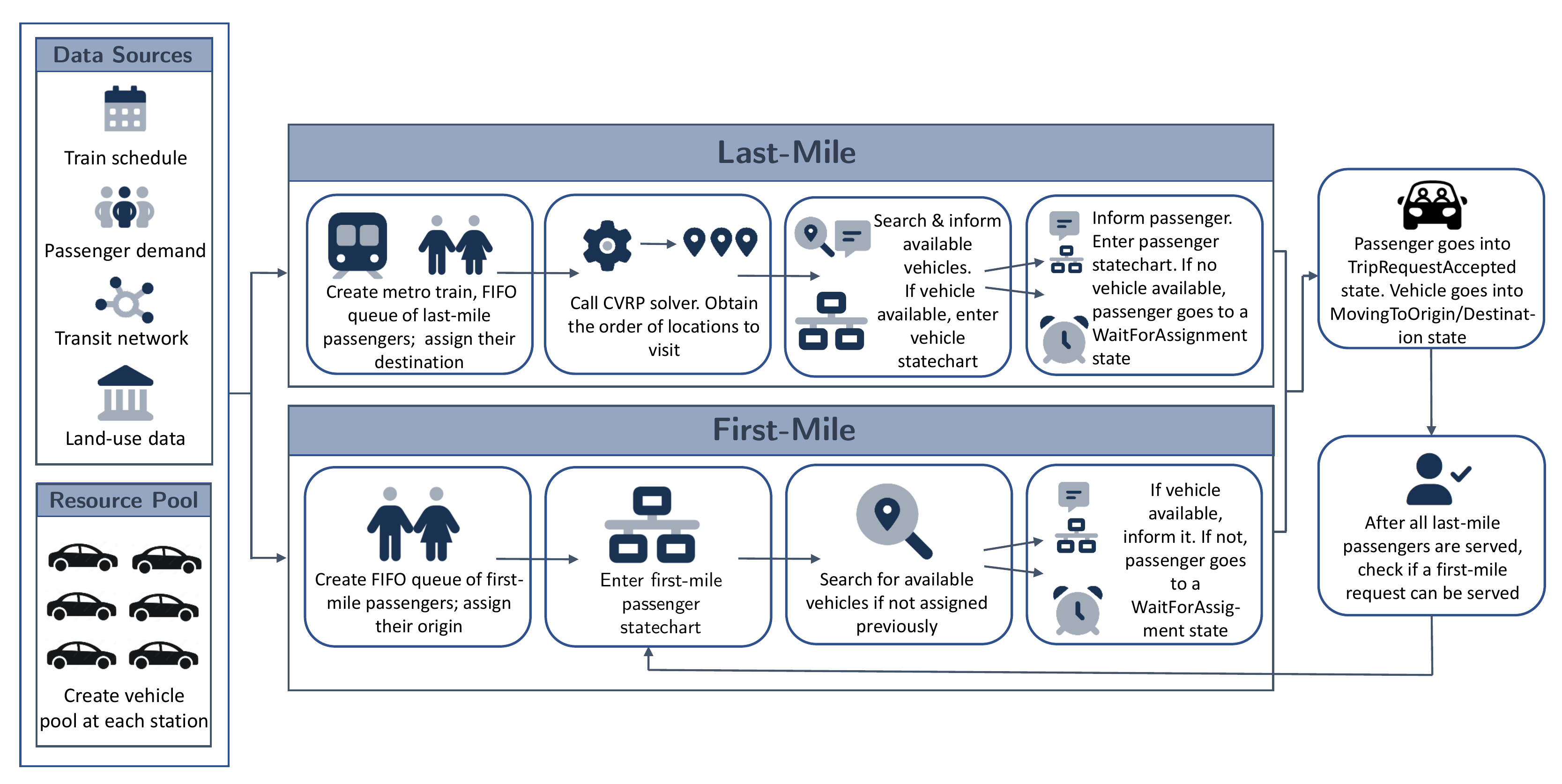}
\caption{Process Overview}
\label{fig:process_overview}
\end{figure}

We now describe the \texttt{Vehicle} and \texttt{LastMilePassenger} agents in detail, along with their statecharts. The statechart for first-mile passengers is modelled similarly.

\begin{table}[t]
    \centering
    \caption{Model agents}
    \begin{tabular}{p{4cm}p{10cm}}
        \hline
         \textit{Agent} & \textit{Description} \\
         \hline
          \texttt{Main} & Top-level agent where other agents reside \\
         \multirow{2}{*}{}\texttt{Vehicle} & Contains information on vehicle fleet \\ & such as origin station and Voronoi region\\
         \multirow{2}{*}{}\texttt{LastMilePassengers} & Contains properties of last-mile passengers \\ & such as their last-mile origin station and destination \\
         \multirow{2}{*}{}\texttt{FirstMilePassengers} & Contains properties of first-mile passengers \\ & such as their origin and first-mile destination station \\
         \multirow{2}{*}{}\texttt{MetroStation} & Stores metro station information including latitude and longitude \\
         \multirow{2}{*}{}\texttt{MetroTrain} & Contains code to trigger events \texttt{eventTrainArrival} \\ & and \texttt{eventPassengerArrival}\\
         \texttt{PassengerDestination} & Stores latitude and longitude of last-mile passenger destinations \\
         \texttt{FirstMileOrigin} & Stores latitude and longitude of first-mile passenger origins \\
         \multirow{2}{*}{}\texttt{LastMileTrip} & Contains last-mile trip information such as \\ & passenger's last-mile origin station and destination \\
         \multirow{2}{*}{}\texttt{FirstMileTrip} & Contains first-mile trip information such as \\ &  passenger's origin and first-mile destination station\\
         \hline
    \end{tabular}
    \label{tab:model_agents}
\end{table}

\begin{figure}[H]
\centering
\includegraphics[scale=0.78]{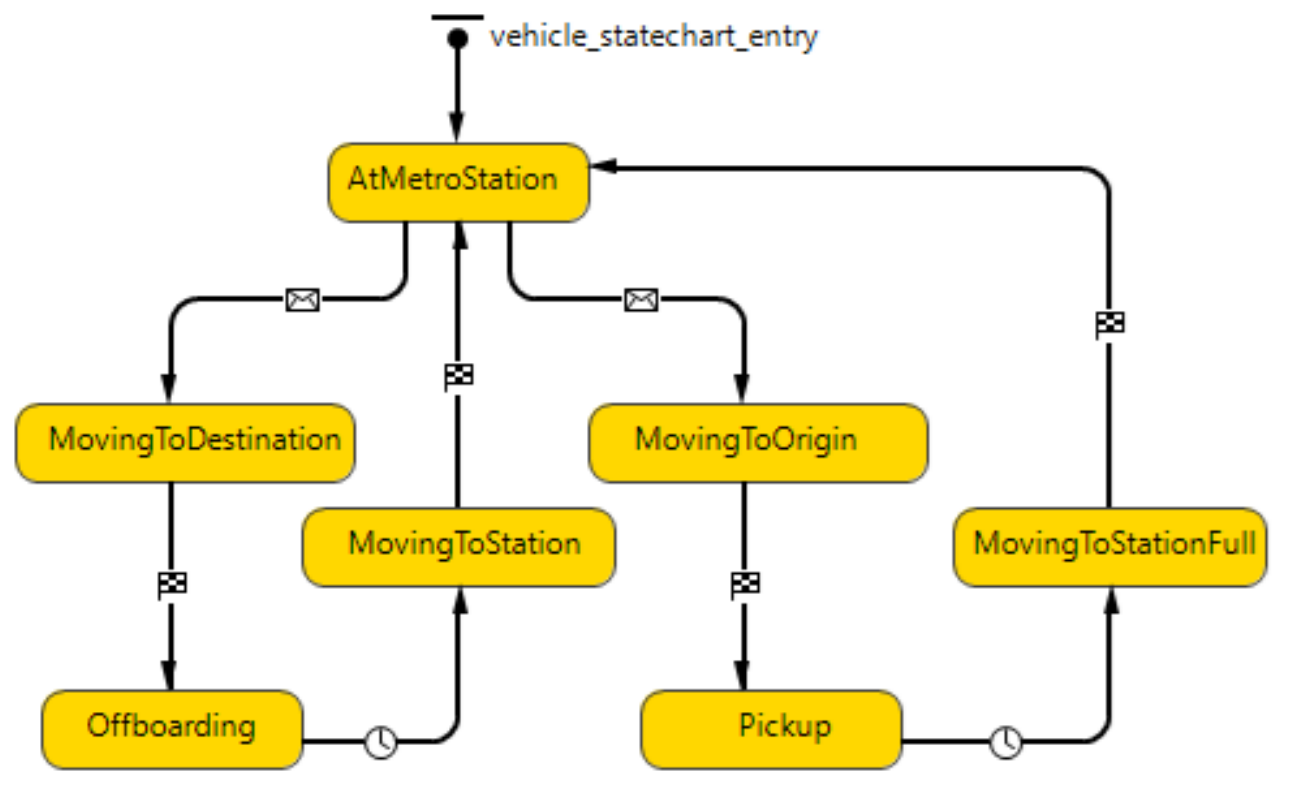}
\caption{FLM vehicle statechart for the no-sharing scenario}
\label{fig:auto_state_chart}
\end{figure}

\textbf{Vehicle agent}:     
Figure \ref{fig:auto_state_chart} describes the statechart of a \texttt{Vehicle} agent. Initially, all FLM vehicle agents are positioned at the metro stations. We first describe the last-mile scenario. A state transition from \texttt{AtMetroStation} state to \texttt{MovingToDestination} happens when the FLM vehicle agent receives a message of the type \texttt{LastMileTrip} that stores information on the last-mile passenger, including their origin station and destination. After receiving this information, the FLM vehicle agent moves to the corresponding destination along the OSM road network, and the vehicle is in the \texttt{MovingToDestination} state. A transition to the \texttt{Offboarding} state occurs by the transition type \texttt{agent\_arrival}, which signifies that the FLM vehicle agent has reached the passenger's last-mile destination. Following a timeout, the FLM vehicle agent moves back to the station in the \texttt{MovingToStation} state. State transition from \texttt{MovingToStation} to \texttt{AtMetroStation} occurs via \texttt{agent\_arrival} transition type.
    
For the first-mile scenario, state transitions take place from \texttt{AtMetroStation} state to \texttt{MovingToOrigin} when a FLM vehicle receives a message of the type \texttt{FirstMileTrip}. On arriving at the origin of a first-mile passenger, the FLM vehicle goes into the \texttt{Pickup} state. Once the first-mile passenger has been picked up, the FLM vehicle agent transitions to the \texttt{MovingToStationFull} state and then back to the \texttt{AtMetroStation} state.

\begin{figure}[h]
\centering
\includegraphics[scale=0.62]{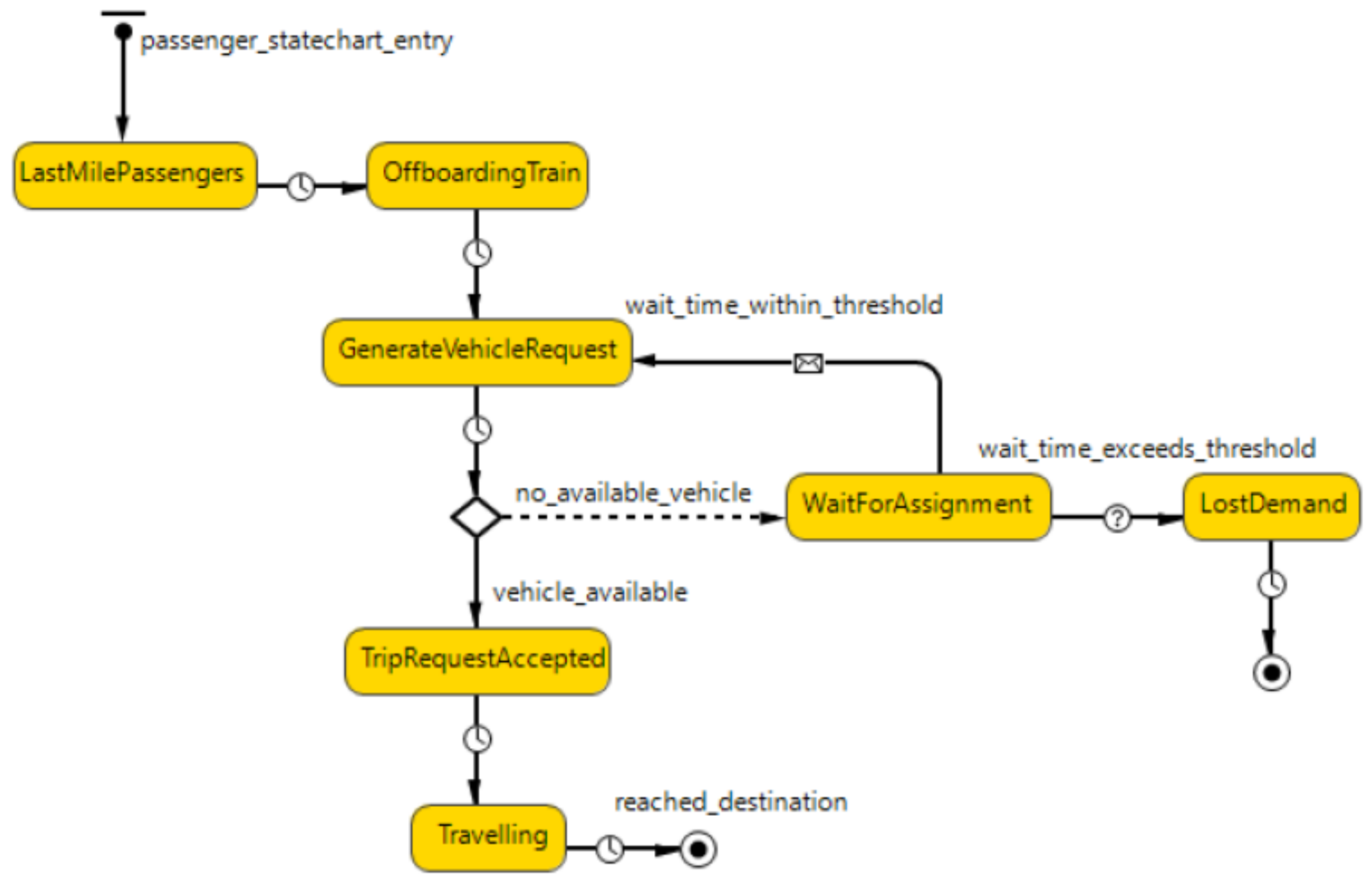}
\caption{Last-mile passenger statechart}
\label{fig:passenger_state_chart}
\end{figure}

\textbf{Passenger agent}: Last-mile passengers alighting at metro stations are modelled as \texttt{LastMilePassenger} agents. Their behaviour is captured using the passenger statechart shown in Figure \ref{fig:passenger_state_chart}. After alighting from the train, each last-mile passenger requests an FLM vehicle in the \texttt{GenerateVehicleRequest} state, and the associated timestamp is recorded. We also maintain a station-wise FIFO queue of passengers containing those who request an FLM service. A decision branch checks the availability of FLM vehicles at the metro station.   

If a vehicle is available, the passenger enters the \texttt{TripRequestAccepted} state. In this state, the passenger is removed from the request queue, and the event \texttt{eventTrip} gets triggered. \texttt{eventTrip} creates an instance of a \texttt{LastMileTrip} agent, assigns vehicles to passengers in a FIFO manner, and sends a message containing the passenger origin station and destination information to the \texttt{Vehicle} agent. The \texttt{Vehicle} agent receives this message through its connections, and the message is then forwarded to its statechart. The passenger then goes into the \texttt{Travelling} state until their destination is reached.  

If an FLM vehicle is unavailable, the passenger goes into the \texttt{WaitForAssignment} state. If the total waiting time exceeds a threshold parameter \texttt{maxWaitingTime}, it is marked as lost demand, and the passenger is removed from the request queue. Otherwise, they retry requesting a vehicle and go to the \texttt{GenerateVehicleRequest} state again, and the process is repeated.
    
\subsection{Simulating Shared Rides}
The scenario described thus far where rides are not shared is considered the base case. Additionally, we model and simulate situations where passenger requests can be pooled together, and an FLM vehicle can serve multiple passengers in a single trip. The following subsections describe these scenarios in detail.

\subsubsection{Last-mile Shareability Case}
\label{subsec:lm_shareability}
\begin{figure}[t]
    \begin{minipage} {0.50\columnwidth}
        \centering
        \subcaptionbox{\small Last-mile shareability}
        {\includegraphics[scale=0.62]{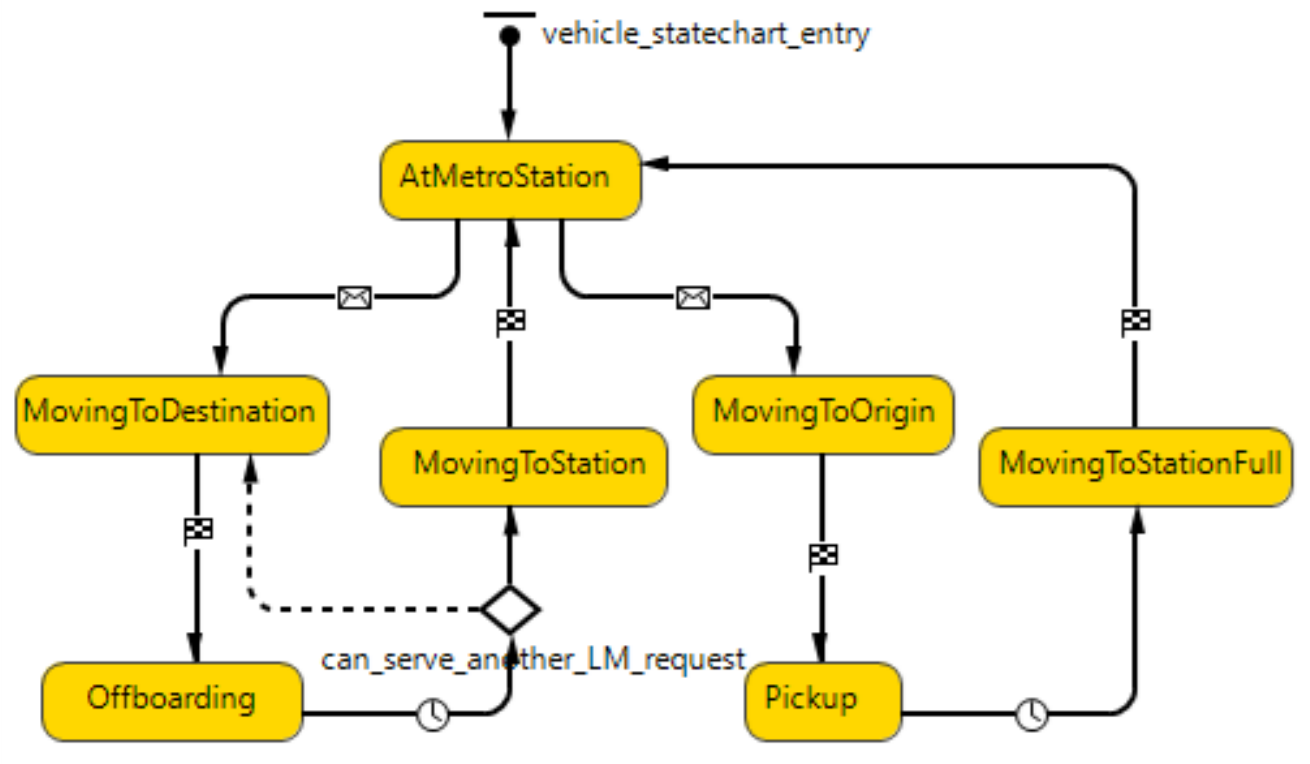}}
        \label{fig:lm_share_auto_state_chart}
    \end{minipage}
    \begin{minipage} {0.50\columnwidth}
        \centering
        \subcaptionbox{\small First-mile shareability}
        {\includegraphics[scale=0.62]{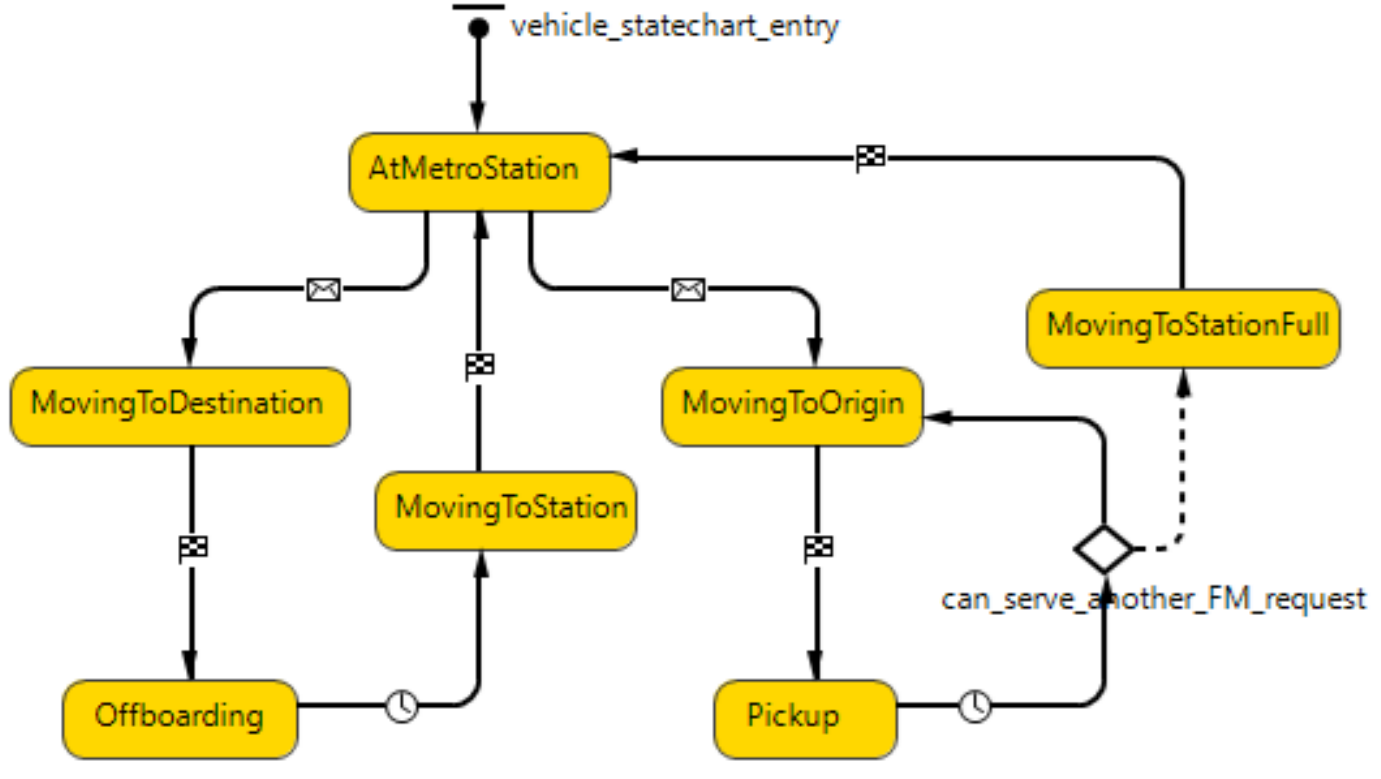}}
        \label{fig:fm_share_auto_state_chart}
    \end{minipage}
    \caption{FLM vehicle statechart for the shareability scenario}
    \label{fig:auto_state_chart_share}
\end{figure}

For the last-mile shareability problem, we pool passengers offboarding at a station. Since multiple passengers arrive at the station simultaneously, we have a sufficient lead time for optimisation and can batch-process their requests. To determine which passengers are to be served in a single trip and the order in which they should be dropped off, we formulate a Capacitated Vehicle Routing Problem (CVRP) \citep{toth2002models} and solve it using Google's OR-Tools \citep{Google_OR-Tools}. The capacity of each FLM vehicle is set to three. The left panel of Figure \ref{fig:auto_state_chart_share} shows the FLM vehicle statechart used. After a last-mile passenger has been offboarded, the control goes from the \texttt{Offboarding} state to the decision branch \texttt{can\_serve\_another\_LM\_request}. Based on the CVRP solution, if there are more passengers to be offboarded, the FLM vehicle goes to the \texttt{MovingToDestination} state of the next passenger. Else, the vehicle goes to the \texttt{MovingToStation} state. In this scenario, we disallow sharing of first-mile trips. In other words, FLM vehicles making a first-mile trip start from the station, pick up a first-mile passenger, and drop them off back at the station.

\subsubsection{First-mile Shareability Case}
\label{subsec:fm_shareability}
In the first-mile shareability scenario, we allow multiple first-mile passengers to be served by the same FLM vehicle in a single trip, while last-mile requests are served one at a time. The right panel of Figure \ref{fig:auto_state_chart_share} shows the FLM vehicle statechart. After a first-mile passenger is picked up by the FLM vehicle in the state \texttt{Pickup}, the control goes to a decision branch \texttt{can\_serve\_another\_FM\_request}. Here, we check the first-mile request queue and pick the passenger whose waiting time is within a threshold. We also ensure that the detour time incurred by the passengers currently on board due to picking up subsequent passengers does not exceed a pre-defined parameter called \texttt{maxDetourTime} threshold. If such a request exists, the FLM vehicle goes to the \texttt{MovingToOrigin} state; else it goes to the \texttt{MovingToStationFull} state. In our experiments, an FLM vehicle is assumed to serve up to three first-mile passengers in a single trip. 

\subsubsection{Joint FLM Shareability Case}
\begin{figure}[H]
    \centering
    \includegraphics[scale=0.78]{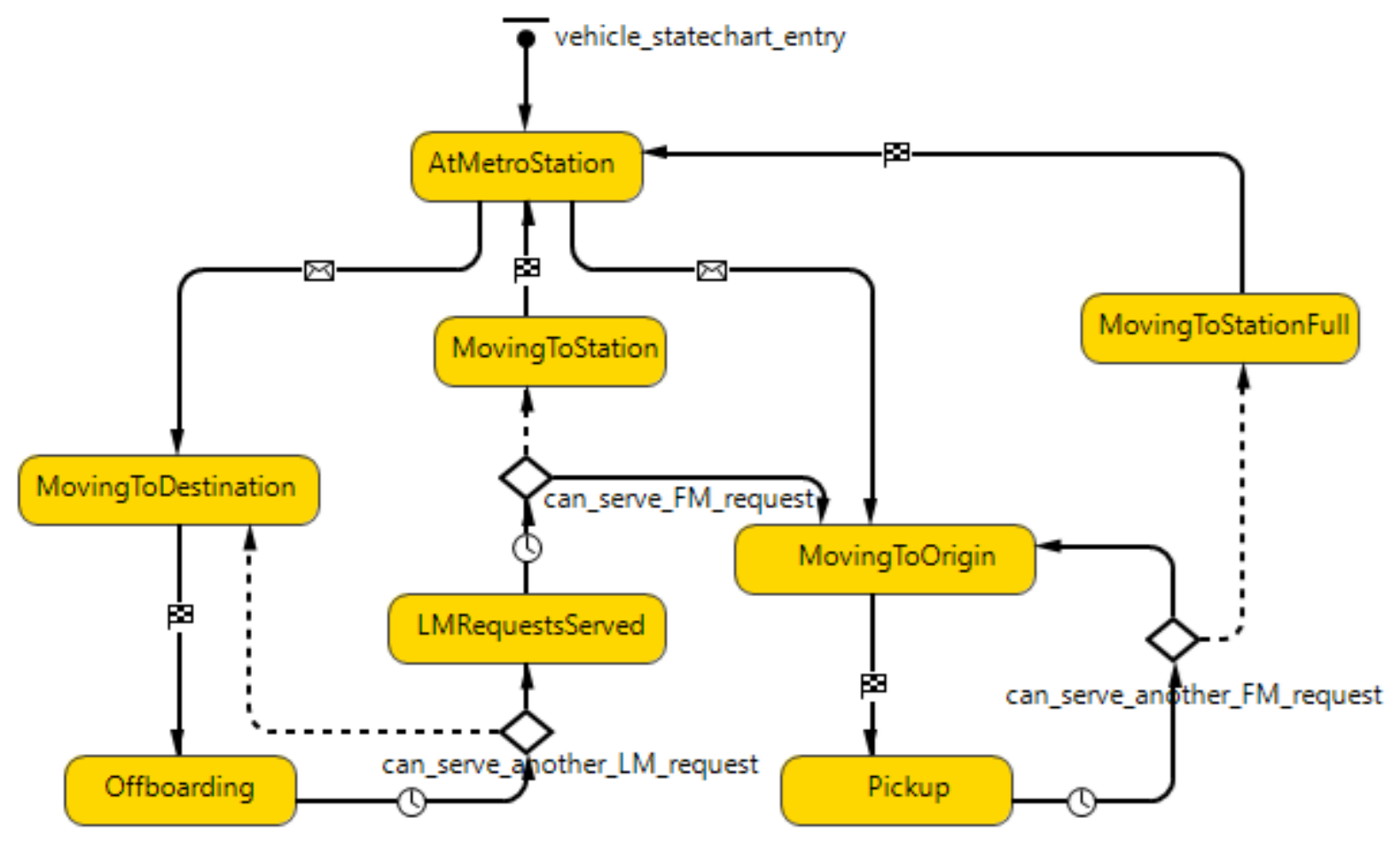}
    \caption{Joint FLM shareability vehicle statechart}
    \label{fig:fm_lm_share_auto_state_chart}
\end{figure}

The joint FLM shareability case combines the scenarios mentioned in Subsections \ref{subsec:lm_shareability}  and \ref{subsec:fm_shareability}. The FLM vehicle statechart shown in Figure \ref{fig:fm_lm_share_auto_state_chart} describes this case. After all last-mile requests are served, the FLM vehicle goes to the \texttt{LMRequestsServed} state. Instead of heading back to the station empty, we introduce a decision branch \texttt{can\_serve\_FM\_request} that checks for first-mile requests that can be served, provided the waiting time is within the threshold. If such a request exists, the vehicle goes to the \texttt{MovingToOrigin} state; else it heads to the \texttt{MovingToStation} state.

\subsection{Integer Linear Programming Model}
\label{subsec:opt_model}
We develop an optimisation model for the no-sharing case using the \textit{Custom Experiment} functionality in AnyLogic and implement it using its Java API and the OptQuest optimisation engine. Custom Experiment offers flexibility for setting up model parameters, seed value, solver run time, and simulation run properties such as start and stop time. During a simulation-based optimisation experiment, multiple simulation runs are carried out by varying the values of decision variables while respecting the specified constraints. For reproducible model runs, it is necessary to reset the seed to a fixed value at the beginning of every iteration. 

The objective in our model was to minimise the total expected lost demand $l_s$ as shown in Equation \eqref{eq:1}. Here, $S$ denotes the set of metro stations, and $s$ represents the index of a station. The decision variables are the number of FLM vehicles assigned to each station, represented by $x_{s}$. We assume a fixed fleet size, and hence the total number of FLM vehicles cannot exceed $a^{total}$ as shown in \eqref{eq:2}. Equation \eqref{eq:3} specifies the upper and lower bounds on the number of vehicles positioned at individual stations, denoted by $a_s^{min}$ and $a_s^{max}$, respectively. Such bounds could be imposed for reasons associated with parking and equity.
\begin{align}
     \min \quad & \sum_{s \in S} l_s(x_s) \label{eq:1} \\
    \text{s.t.\quad} & \sum_{s \in S} {x}_{s} \leq a^{total} \label{eq:2} \\
    & a_s^{min} \leq {x}_{s} \leq a_s^{max} & & \forall \, \, s \in S \label{eq:3} \\
    & x_s \in \mathbb{Z}_{+} && \forall \, \, s \in S
\end{align}
We observed that the time taken by OptQuest to complete a single iteration increases substantially in the last-mile and joint shareability case due to function calls to OR-Tools for solving the CVRP. However, note that the objective function can be decomposed by stations, and only Constraint \eqref{eq:2} binds the decision variables. It is expected that as the number of FLM vehicles assigned to a station increases, the lost demand will decrease. Thus, this relationship is likely to be decreasing (and in most cases convex, with diminishing returns for every extra vehicle). 

\begin{figure}[H]
    \centering
    \includegraphics[scale=0.45]{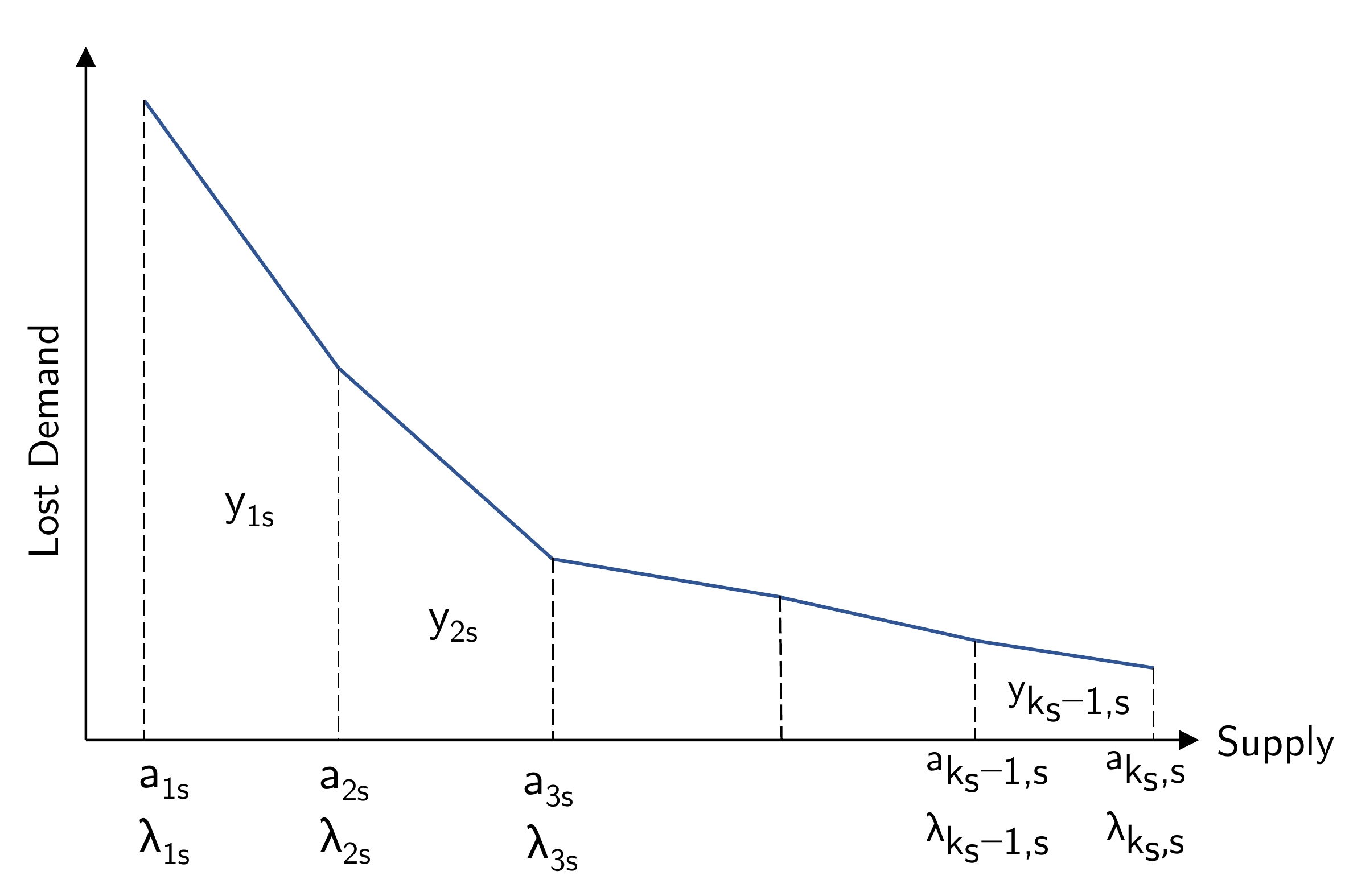}
    \caption{Approximation of the lost demand curve for a metro station $s$}
    \label{fig:piecewise}
\end{figure}

At each station, we interpolate the expected lost demand using simulation runs performed at uniformly discretised supply values (or break points) as shown in Figure~\ref{fig:piecewise}. With such lost demand curves for each station as problem input, we propose a piecewise linear cost approximation for the objective and formulate it as an Integer Linear Program (ILP), which is solved using CPLEX. The ILP model uses auxiliary variables and has an objective that is separable with respect to the supply at stations.

Let the piecewise linear function of station $s$ be described using function values $l_s(a_{is})$, where $a_{is}$ is the $i^{th}$ break point at which the simulation is performed. Let the number of pieces in the lost demand curve for station $s$ be $k_s-1$. Define $\lambda_{is}$ as the weight placed on the break point $a_{is}$. Let $y_{is}$ be one if the optimal solution for station $s$ lies in piece $i$.    
\begin{align}
   \min & \, \,  \sum_{s \in S} \sum_{i = 1}^{k_s} \lambda_{is} l_{is}(a_{is})  \label{eq:obj} \\
     \text{s.t.} &  \, \, \sum_{s \in S} \sum_{i = 1}^{k_s} \lambda_{is} a_{is} \leq a^{total} \label{eq:supply_limit} \\
     & \, \,  \sum_{i=1}^{k_s-1} y_{is} = 1 &  \forall s \in S \label{eq:ys} \\
         & \, \,  \sum_{i = 1}^{k_s} \lambda_i = 1 & \forall s \in S  \label{eq:lambdas} \\
     & \, \, \lambda_{1s} \leq y_{1s} & \forall s \in S \label{eq:ylambda1} \\
    & \, \, \lambda_{is} \leq y_{i-1,s} + y_{is} & \forall s \in S, i = 2, \ldots, k_s-1 \label{eq:ylambda2}\\
    & \, \, \lambda_{k_s,s} \leq y_{k_s-1,s} &  \forall s \in S \label{eq:ylambda3} \\
    & \, \, \lambda_{is}  \geq 0 & \forall s \in S, i = 1, \ldots, k_s \label{eq:bounds1} \\
    & \, \, y_{is}  \in  \{0,1\} & \forall s \in S, i = 1, \ldots, k_s-1 \label{eq:bounds2}
\end{align}

The number of FLM vehicles assigned to a station $s$ is represented as $\sum_{i = 1}^{k_s} \lambda_{is} a_{is}$ and hence Constraint \eqref{eq:supply_limit} ensures that the total supply is not exceeded. The optimal supply for a station $s$ must lie in one of the pieces, which is enforced by \eqref{eq:ys}. The actual supply value is determined by a convex combination of the $\lambda$s using \eqref{eq:lambdas}. If the $y$ variable is 1, only the $\lambda$s associated with its endpoints must take non-negative values. This constraint is enforced by \eqref{eq:ylambda1}--\eqref{eq:ylambda3}. Finally, \eqref{eq:bounds1} and \eqref{eq:bounds2} impose bounds and integrality constraints. We do not require constraints to model \eqref{eq:3} in this formulation because they are implicitly set by $a_{1s}$ and $a_{k_s,s}$.

\begin{figure}[ht]
    \centering
    \includegraphics[scale=0.35]{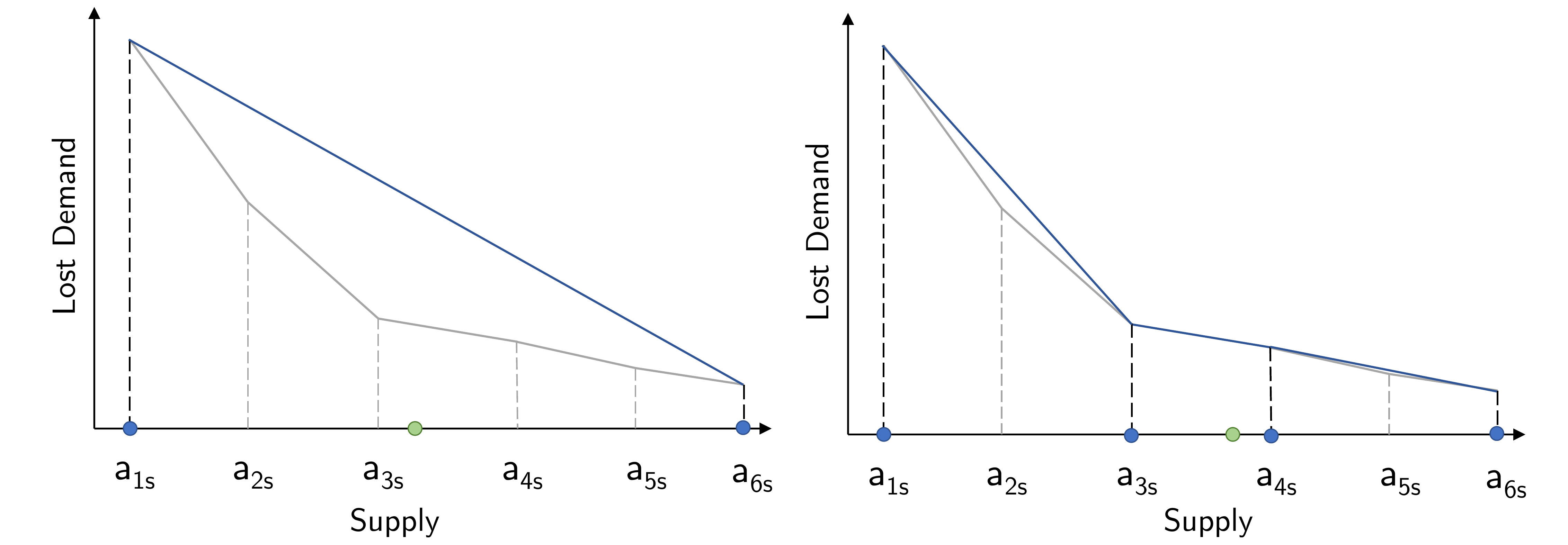}
    \caption{Adaptive refinement of the feasible region}
    \label{fig:adaptive}
\end{figure}

The proposed ILP can be solved reasonably quickly using off-the-shelf solvers. However, the major bottleneck is in estimating the lost demand at the break points using simulation. This problem can be addressed by starting with a coarse set of break points and iteratively generating new ones. To illustrate this idea, consider the example in Figure \ref{fig:adaptive}. Suppose that for a station $s$, the lost demand was estimated at two break points $a_{1s}$ and $a_{6s}$. We approximate the lost demand curve as shown by the thick line segments in the left panel. Suppose the resulting optimal ILP solution lies in the interval $[a_{3s}, a_{4s}]$. Additional simulations are then run to estimate the lost demand at the endpoints $a_{3s}$ and $a_{4s}$, and the lost demand curve is updated with more pieces as shown in the right panel. If both endpoints of the optimal solution are already explored, we stop. If not, we perform new simulations at new break points, update lost demand curves, and repeat. If the ILP solution coincides with the break point, say $a_{3s}$, we explore its neighbours $a_{2s}$ and $a_{4s}$. This adaptive procedure was found to reduce the number of break points and simulations by nearly 50\%.

\section{Experiments and Results}
\label{sec:experiments}

One day of Bengaluru metro operations was simulated using train schedules, which included trips starting between 5 AM--11 PM and passenger boarding and alighting information from January 2018. Based on data on ride-sharing trips in Bengaluru, the threshold for maximum waiting time and detour time was set to 7 minutes \citep{wait_time}. If trip requests are not served within this time, they are counted as lost demand. FLM vehicles are assumed to maintain a uniform speed of 21.2 kmph, the average speed of motorised vehicles in Bengaluru \citep{auto_speed}. The total number of FLM vehicles $a_{total}$ is fixed at 1200, and the lower and upper bounds at each station $s$, $a_s^{min}$ and $a_s^{max}$, are set to 5 and 60, respectively. The Java code from AnyLogic and the CVRP Python functions can be found at \url{github.com/transnetlab/agent-based-fleet-management} for reference.

\subsection{Lost Demand}
We benchmark our proposed method for fleet allocation by comparing it with two baseline strategies and the OptQuest solver.

\textbf{Baseline Strategies}: Two simple strategies that could be used by an operator in the absence of an optimisation model were simulated. In the first strategy (\textit{proportional to demand}), we try to achieve a supply allocation proportional to the total first- and last-mile demand. However, since the number of vehicles allowed at each station is capped at 60, vehicles in excess of this value are again redistributed among the remaining stations, proportional to their corresponding demand and the process is repeated until all 1200 vehicles are assigned. In the second case (\textit{equal supply}), we simply allocate an equal number of FLM vehicles (i.e., 1200/40 = 30) at each metro station. The results for all four shareability scenarios are shown in the first two columns of Table \ref{tab:obj_value}. The values in brackets show percentage benefit compared with the proportional to demand strategy. Assigning an equal number of vehicles across stations resulted in higher lost demand in all shareability scenarios.

\textbf{OptQuest}: The optimal fleet allocation problem was also solved using AnyLogic's in-built OptQuest solver with a time budget of 24 hours. The solver was initialised with 30 vehicles at each station. For the no sharing case, a total of 1235 iterations were simulated, but the best-found solution's objective was 17,854, which was 19.4\% worse compared to the proportional to demand case. The quality of solutions discovered for the shareability scenarios was also poorer compared with the proportional to demand case, mainly because each of these iterations runs slower when sharing is involved. Hence, only a limited portion of the feasible region is explored.

%Path: Shared Folder Visualization/Equal&Proportional_RunResults.xlsx. OptQuest values from Shared Folder Visualization/OptQuest folder
\begin{table}[H]
    \centering
    \caption{Comparison of objective values for no-sharing versus shareability scenarios}
    \begin{tabular}{p{3.5cm}cx{2.4cm}cx{2.5cm}cx{2.5cm}cx{2.5cm}}
        \hline
         \textit{Scenario} & \textit{Proportional to demand} & \textit{Equal supply} & \textit{OptQuest} & %\textit{CPLEX} & 
         ILP Model\\
         \hline
          No sharing & 14,957 & 21,785 (-45.7\%) & 17,854 (-19.4\%) & 14,379 (3.9\%) \\
          Last-mile shareability & 11,539 & 17,265 (-49.6\%) & 17,109  (-48.3\%) & 10,903 (5.5\%) \\
          First-mile shareability & 9,597 & 16,783 (-74.9\%) & 16,375 (-70.6\%) & 8,821 (8.1\%) \\
          Joint FLM shareability & 3,655 & 7,738 (-111.7\%) & 7779 (-112.8\%) & 2,714 (25.7\%)\\
         \hline
    \end{tabular}
    \label{tab:obj_value}
\end{table}

\begin{figure}[H]
    \centering
    \includegraphics[scale=0.45]{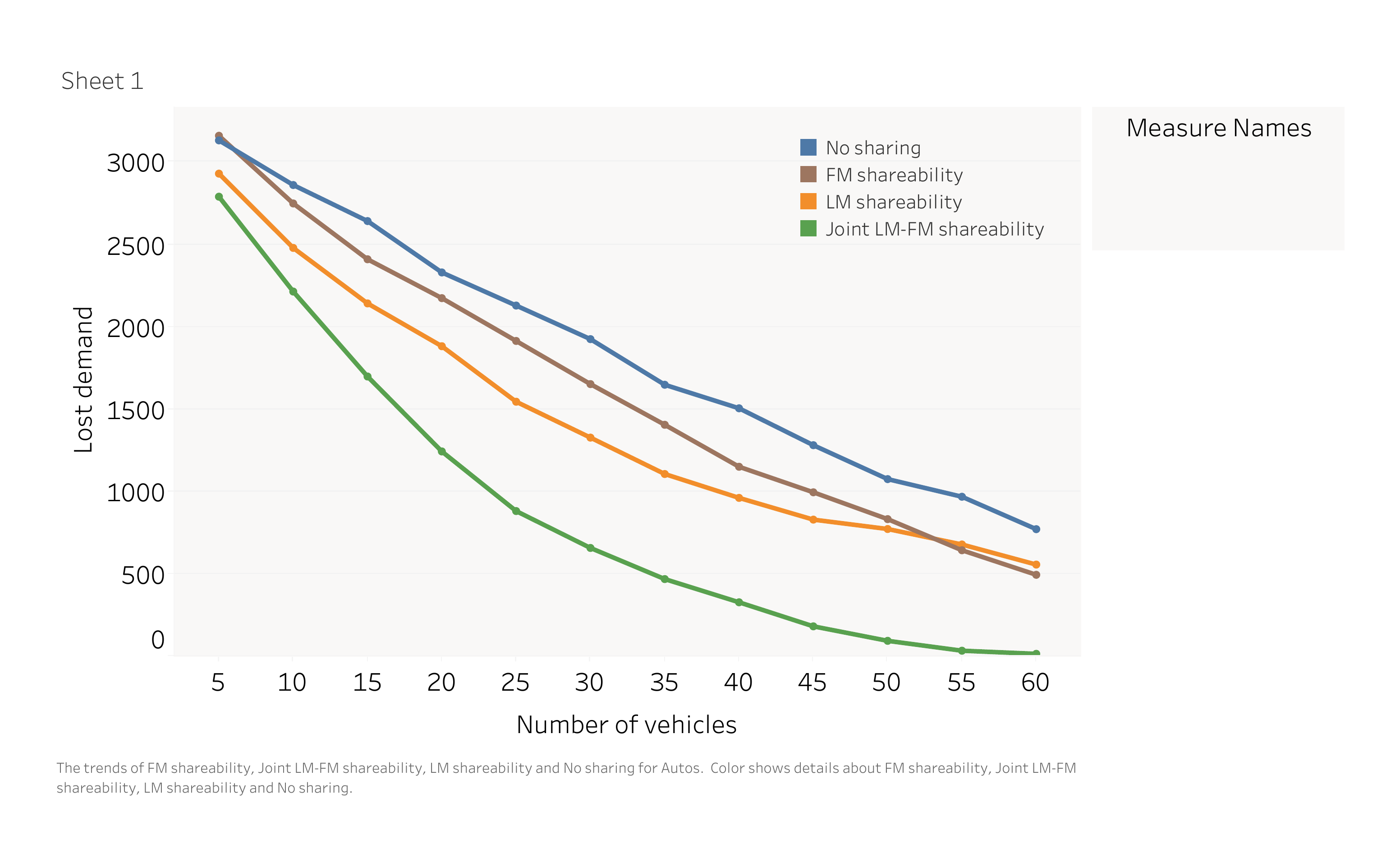}
    \caption{Lost demand comparison for a metro station -- Indiranagar -- under different shareability scenarios }
    \label{fig:compare_share_vs_noShare}
\end{figure}

\textbf{ILP Model}: As described in Section \ref{subsec:opt_model}, we discretised the feasible region and ran simulations by varying the number of vehicles at each station from 5 to 60 in increments of 5 and estimated the expected lost demand. For the adaptive piecewise approximation method, only 247 out of 480 simulations (40 stations with 12 break points at each station) were needed to find the optimum supply values. Figure \ref{fig:compare_share_vs_noShare} compares the lost demand for different shareability scenarios for one of the metro stations -- Indiranagar. The joint FLM shareability case results in significantly lower lost demand for all supply values. Similar trends were observed at other stations. The average number of simulated first- and last-mile passengers in this scenario was 36,664 and 36,994, respectively. 

These curves were used as input to the ILP formulation \eqref{eq:obj}--\eqref{eq:bounds2}, which was solved using CPLEX. The optimal decision variables, i.e., the number of FLM vehicles at every station, were passed to the AnyLogic simulation, the results of which are shown in the last column of Table \ref{tab:obj_value}. The CPLEX runtimes in all four shareability scenarios were less than a minute. The estimated lost demand for the baseline strategies and ILP model are average values from five simulation runs. 

%Path: Shared Folder Visualization/lost_demand_Indiranagar_Tableau.xlsx

\vspace*{-6mm}

%Path: Shared Folder Visualization/Lost_demand_heatmap.xlsx
\begin{figure}[H]
    \centering
    \includegraphics[scale=0.146]{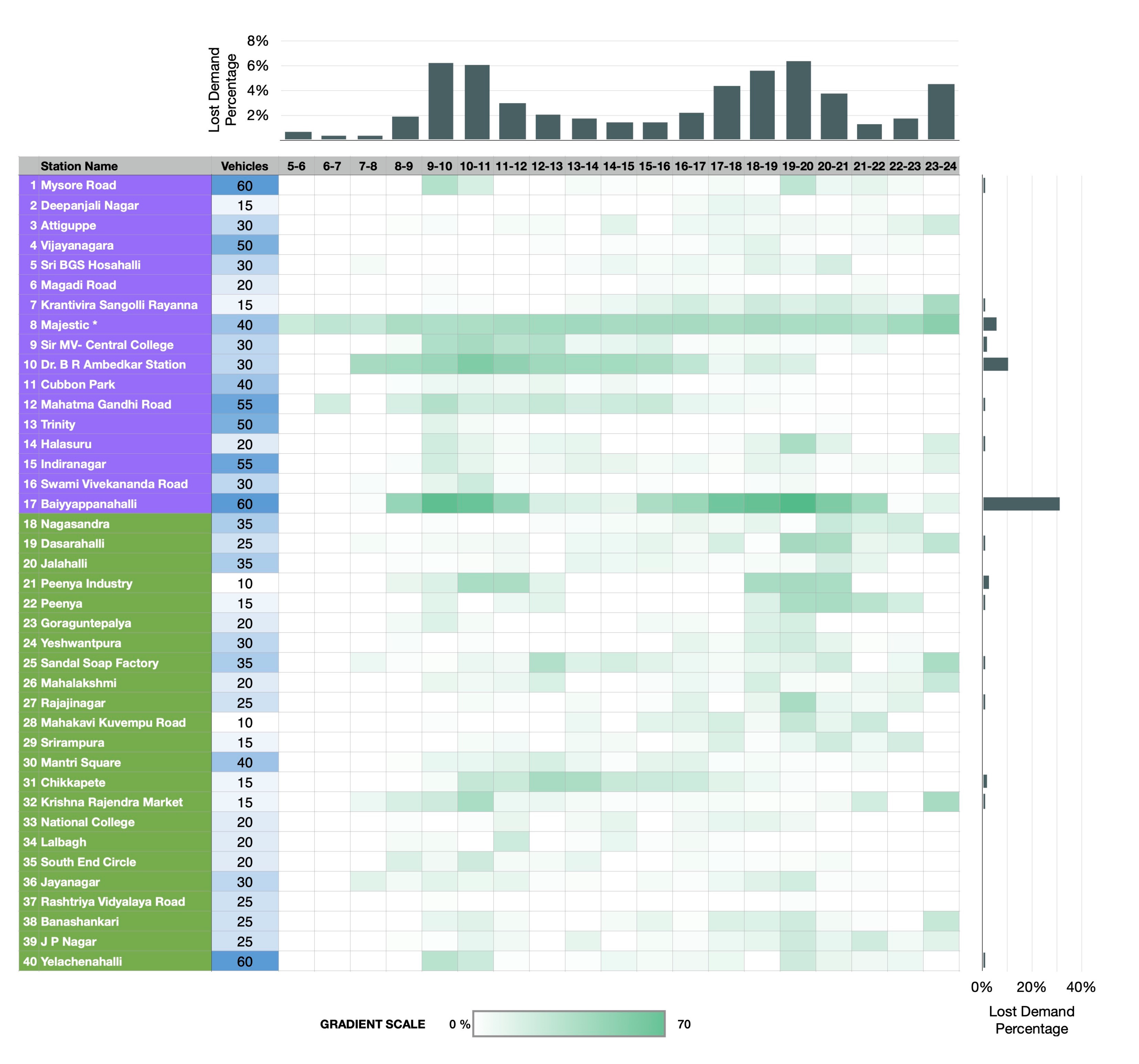}
    \caption{Spatio-temporal distribution of lost demand}
    \label{fig:joint_plot}
\end{figure}

Overall, the ILP model solution consistently resulted in lower lost demand and outperformed the other allocation methods for all shareability scenarios. Specifically, for the joint FLM shareability case, improvements of 65\% and 26\% were observed over the baseline strategies, showcasing the potential of our work in improving FLM connectivity. The rest of this section analyses the solutions from the proposed ILP method for the joint FLM shareability scenario in greater detail.

\begin{figure}[H]
    \centering
    \includegraphics[scale=0.6]{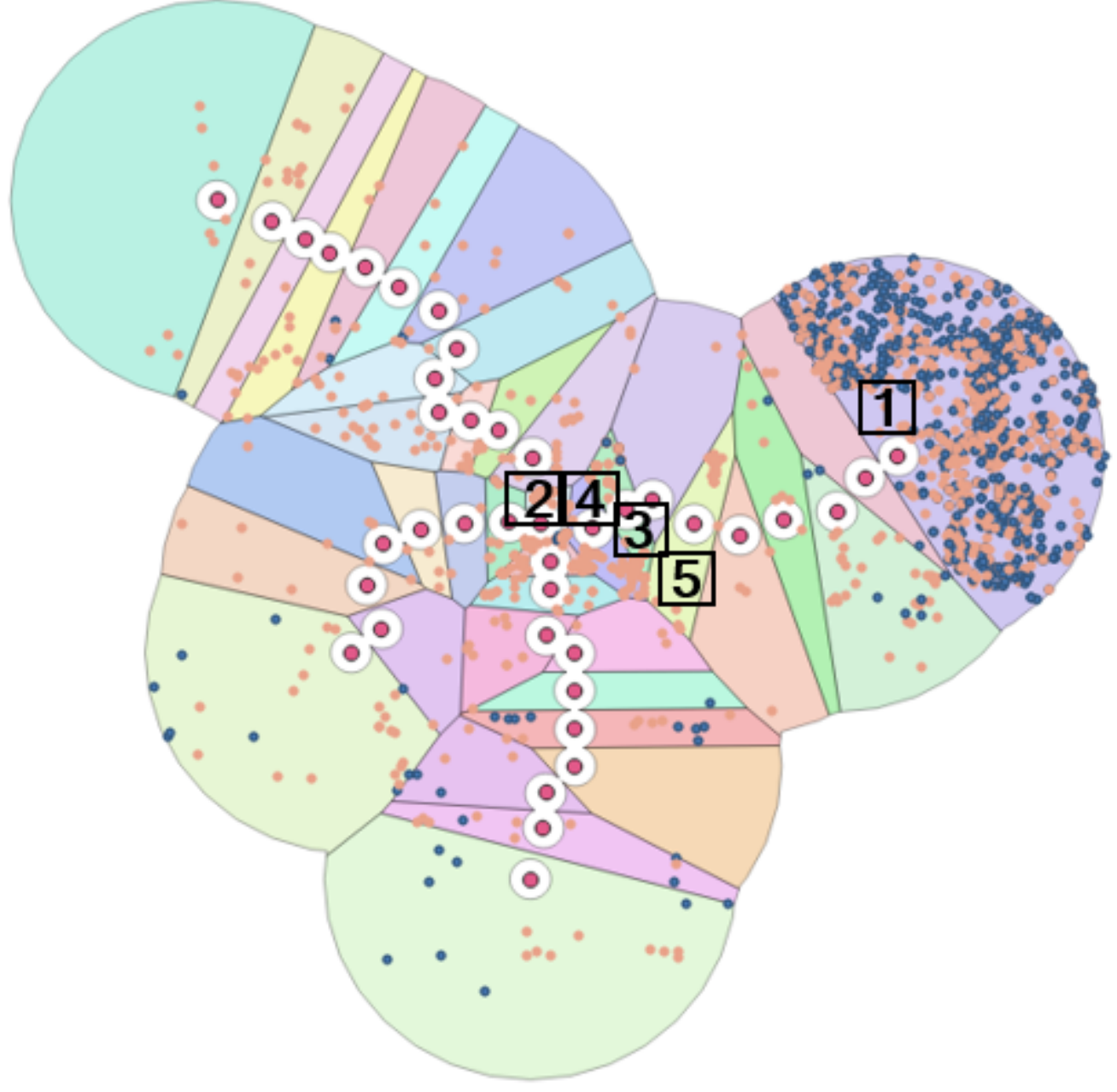}
    \caption{Locations of lost demand for a single simulation run. Last-mile lost demand locations are shown in orange, while blue dots denote first-mile lost demand locations.}
    \label{fig:lost_demand_location}
\end{figure}

The optimal number of vehicles at each station and the lost demand percentage, calculated with respect to the total demand on an hourly and station-wise basis, are shown in Figure \ref{fig:joint_plot}. The bar plots on the top and right indicate the cumulative lost demand percentages for each hour and each station, respectively. Lost demand is highest in the time window 7 PM--8 PM, followed by 9 AM--10 AM. As expected, the off-peak morning hours 5 AM--8 AM record the lowest lost demand percentage. For most metro stations, the percentage of lost demand is less than 5\%. Three out of four ends of the metro lines are assigned the maximum number of FLM vehicles allowed because of their large Voronoi regions. Among them, Baiyyappanahalli (label \#17), one of the endpoints on the East-West line, has the highest lost demand (see Figure \ref{fig:lost_demand_location}, which shows the locations of the lost demand). As mentioned earlier, Nagasandra station in the North also has a sizeable Voronoi region, but the complete EB data was unavailable. The lost demand at the other two terminal stations -- Mysore Road and Yelachenahalli -- is relatively much lower because the passenger demand at these stations is nearly half that at Baiyyappanahalli. The overall difference between the number of last- and first-mile lost demand across all stations for this instance was found to be 413. 

%Path: Shared Folder Visualization/lost_demand_location.csv

Table \ref{tab:highest_five} shows the top five metro stations with the highest lost demand, their LM and FM demand, the number of FLM vehicles in the optimal solution, and LM and FM lost demand. Their rank based on the total lost demand is also indicated in Figure \ref{fig:lost_demand_location}. Despite having a significant share of lost demand, the stations Majestic and Dr. BR Ambedkar Station (labels \# 8 and \#10) are not assigned more FLM vehicles because the marginal effect of adding an extra vehicle at these stations was found to be very low. The number of unserved first-mile requests in the case of Baiyyappanahalli is higher because the detour times are longer compared with other stations with a high passenger volume, such as Majestic.

%Shared Folder Visualization/Tableau_AnyLogic_results.xlsx. Sheet names: "consolidated_results"
\begin{table}[H]
    \centering
    \caption{Top five stations with highest lost demand}
    \begin{tabular}{clccccc}
        \hline
         \textit{S.No.} & \textit{Station} & \textit{LM demand} & \textit{FM demand} & \textit{\# Vehicles} & \textit{LM lost demand} &
         \textit{FM lost demand}\\
         \hline
          1 & Baiyyappanahalli & 2888 & 2750 & 60 & 724 & 1025 \\
          2 & Majestic & 3153 & 3083 & 40 & 350 & 2 \\
          3 & Dr. BR Ambedkar Stn & 897 & 881 & 30 & 182 & 2 \\
          4 & Sir MV Central Clg & 1173 & 1133 & 30 & 48 & 0 \\
          5 & Mahatma Gandhi Rd & 1642 & 1587 & 55 & 35 & 1 \\
         \hline
    \end{tabular}
    \label{tab:highest_five}
\end{table}

To test the sensitivity of the optimal solution with respect to the total number of vehicles, we ran a few additional scenarios. Specifically, we set the number of vehicles to 800, 1200, and 1600. The percentage of lost demand in these instances was 10.7\%, 3.7\%, and 3.5\%, respectively.

\subsection{Vehicle Utilisation}
\begin{figure}[H]
    \centering
    \includegraphics[scale=0.47]{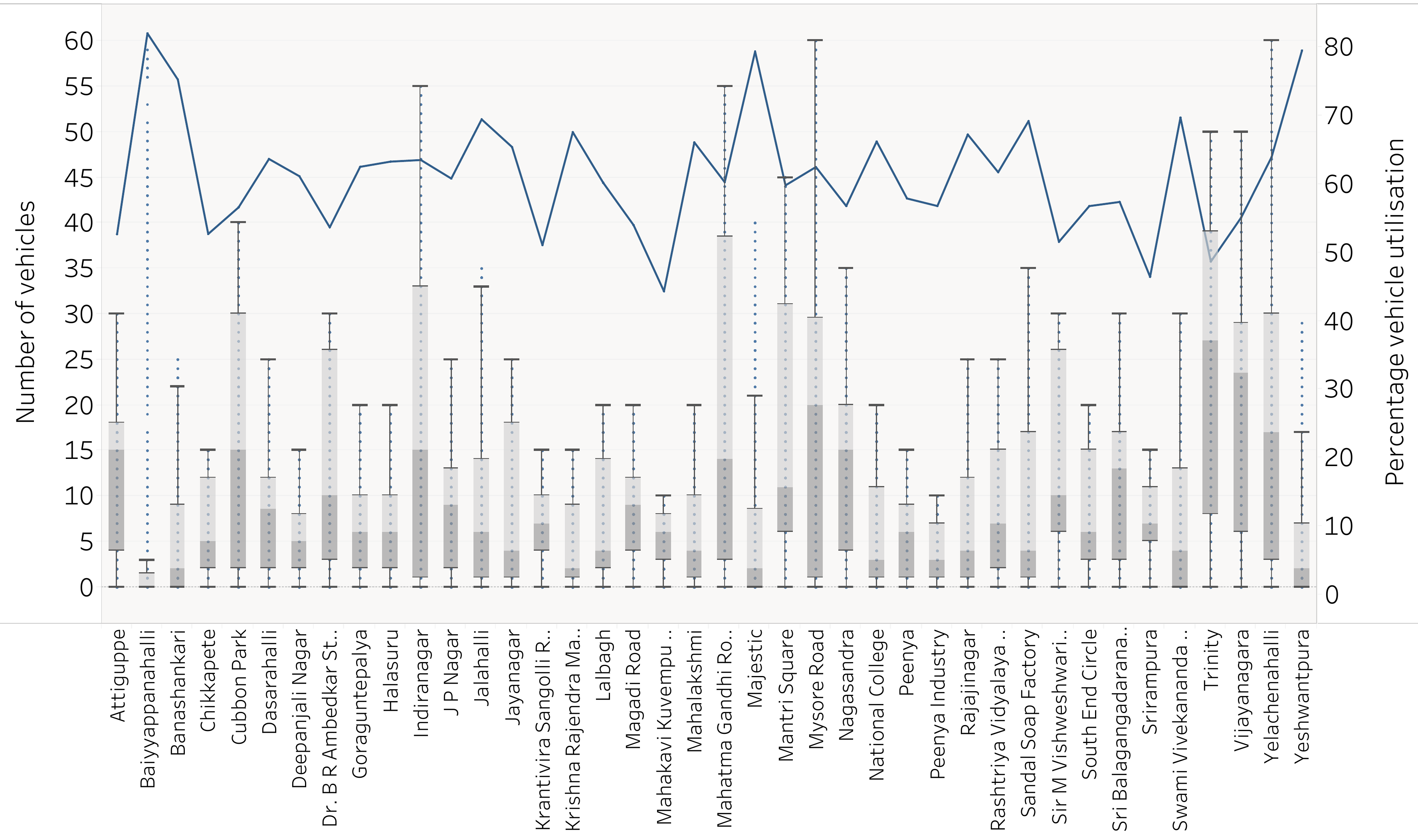}
    \caption{Box plot showing the number of vehicles parked at different stations. The line chart shows the percentage vehicle utilisation.}
    \label{fig:auto_count}
\end{figure}

We also compared utilisation statistics of FLM vehicles in the system. The line chart associated with the secondary axis in Figure \ref{fig:auto_count} shows the average unit utilisation of vehicles assigned to different stations. Unit utilisation values indicate the percentage of time the vehicle was busy. The highest (82\%) and lowest (44\%) utilisation percentages were found to correspond to stations Baiyyappanahalli (label \#17) and Mahakavi Kuvempu Road (label \#28), respectively, and are correlated with the supply and demand at these stations. Knowing utilisation percentages can help reserve downtime of the vehicle fleet if needed (e.g., for maintenance,  refuelling, and recharging/battery swapping in the case of an electric fleet). 

Vehicle utilisation statistics are also a measure of the number of idle vehicles that require parking at stations. Figure \ref{fig:auto_count} shows a box plot in which the number of vehicles parked at different stations estimated at a one-minute frequency makes up the data points. As expected, stations with high utilisation, such as Baiyyappanahalli, Majestic, and Yeshwantpura, have low median values. These percentile measures can be used to design appropriate parking infrastructure. 

%Shared Folder Visualization/Tableau_AnyLogic_results.xlsx; Sheet name: "auto_count_stats (2)"
%Sheet name: "auto_count_stats (2)". Tableau sheet name "Vehicle Parking+Utsn Stats"

\begin{figure}[h]
    \begin{minipage} {0.50\columnwidth}
        \centering
        \subcaptionbox{\small Indiranagar station}
        {\includegraphics[scale=0.47]{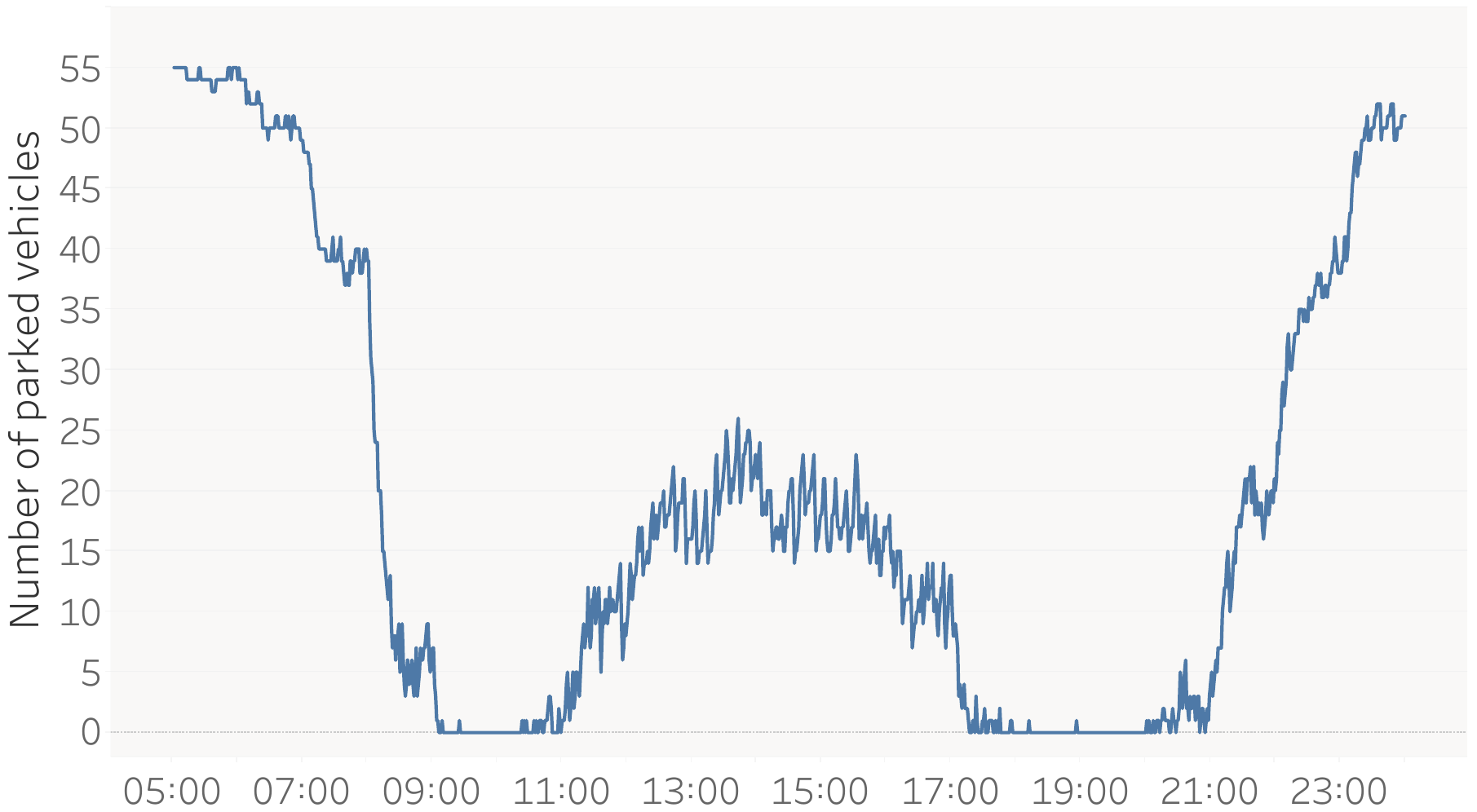}}
        \label{fig:time_series_Indiranagar}
    \end{minipage}
    \begin{minipage} {0.50\columnwidth}
        \centering
        \subcaptionbox{\small Majestic station}
        {\includegraphics[scale=0.47]{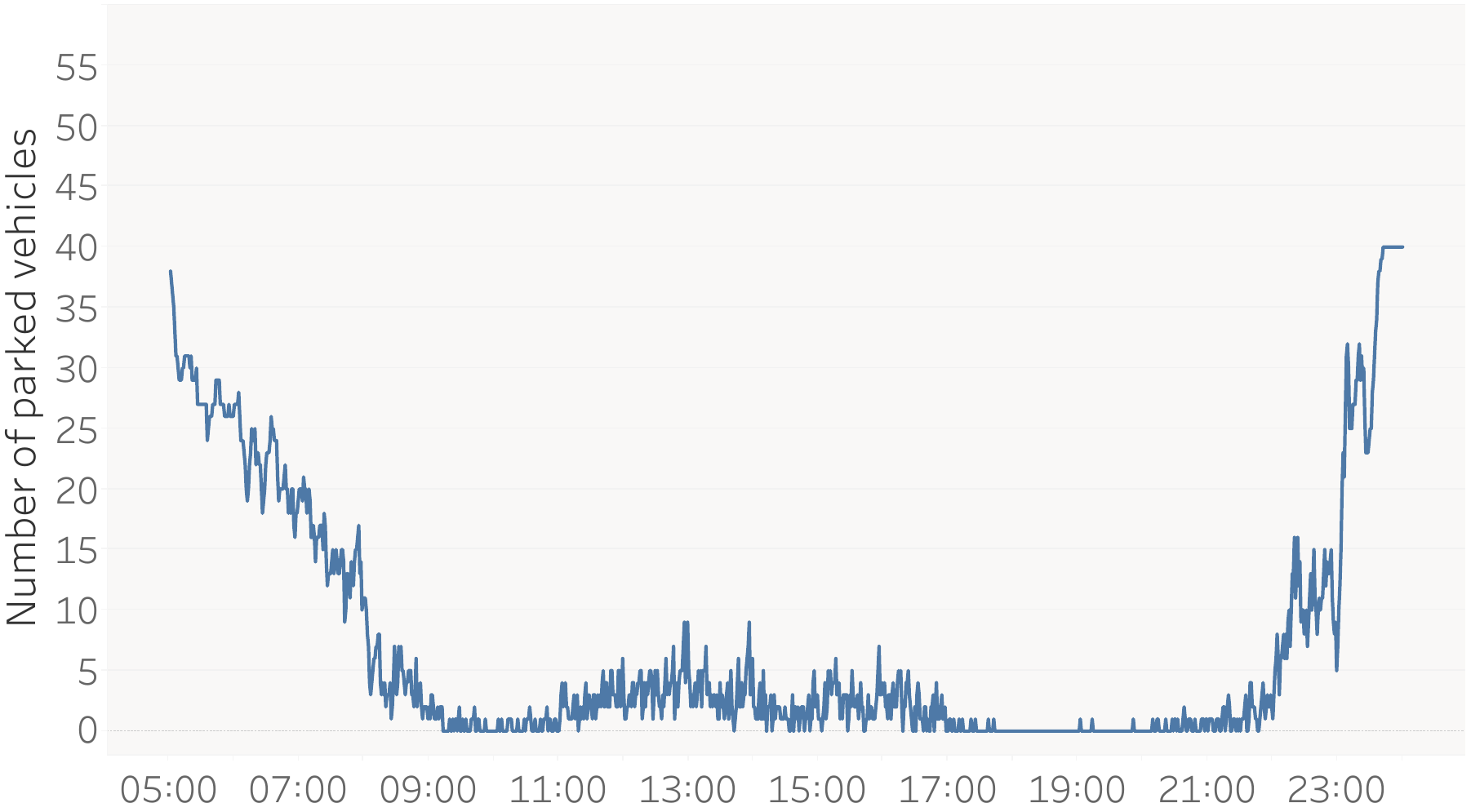}}
        \label{fig:time_series_Majestic}
    \end{minipage}
    \caption{Time series plots showing the number of vehicles parked at the metro stations at different hours of the day.}
    \label{fig:time_series}
\end{figure}

Parking requirements are not usually uniform and mirror the demand distribution. Figure \ref{fig:time_series} shows temporal variations in the number of idle FLM vehicles for two stations -- Indiranagar and Majestic. The parking requirements are lowest during the morning (9 AM--11 AM) and evening peak hours (6 PM--8 PM), but the off-peak requirements can be significantly different across stations, as seen from the figure.  

\subsection{Shareability}
\begin{figure}[H]
    \centering
    \includegraphics[scale=0.5]{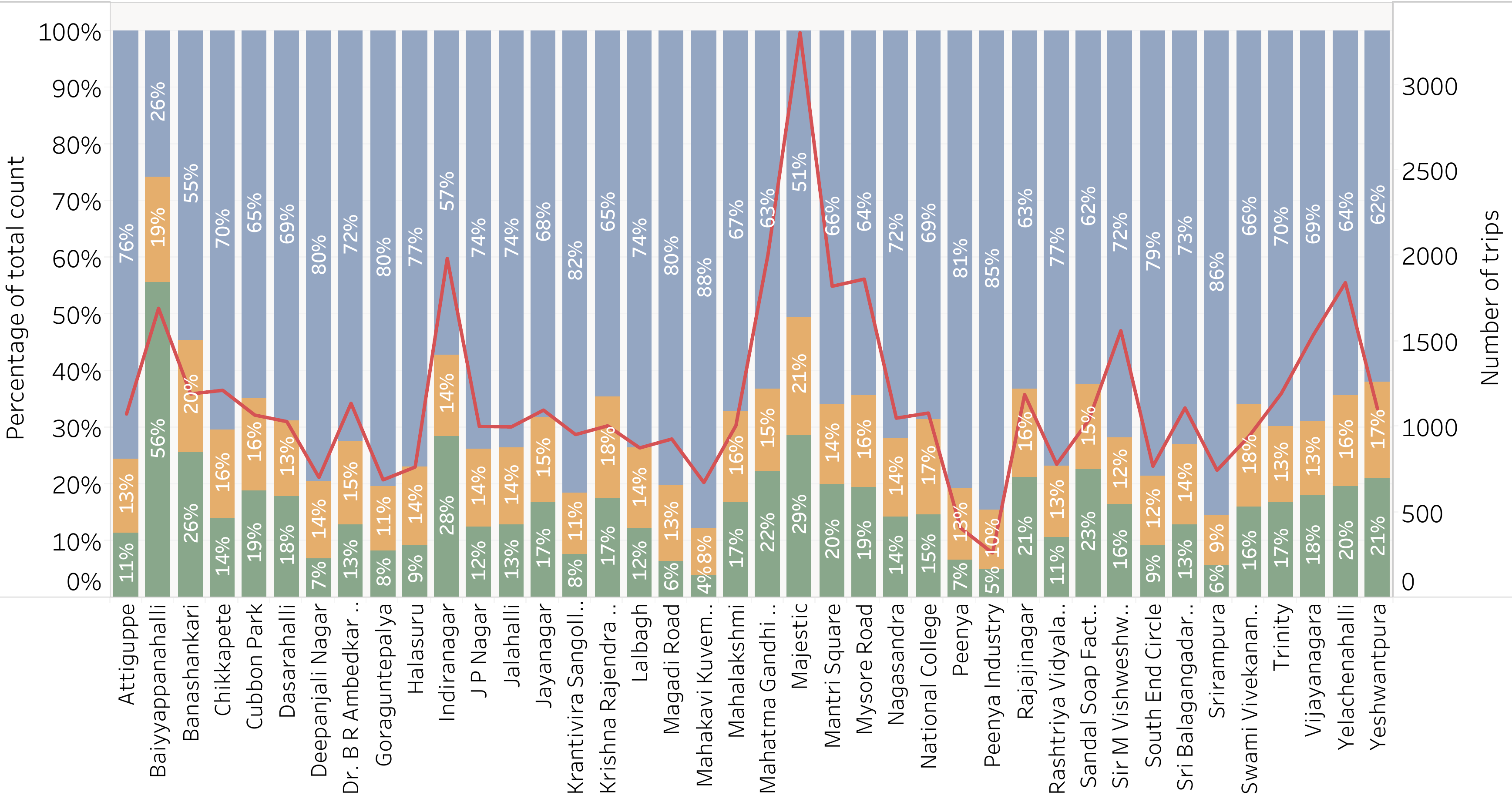}
    \caption{Percentage of trips with 1, 2, and 3 passengers. Single-passenger, 2-sharing, and 3-sharing trips are represented by blue, yellow, and green bars, respectively. The line plot shows the average number of trips for each station.}
    \label{fig:shareability_stats}
\end{figure}

Figure \ref{fig:shareability_stats} shows the percentage of trips that involve transporting a single passenger and ride-sharing with two or three passengers on board. The superimposed line plot indicates the total number of trips. Recall that in the joint shareability model (see Figure \ref{fig:fm_lm_share_auto_state_chart}), FLM vehicles first serve last-mile demand and switch to picking up first-mile passengers (if available and if waiting/detour time thresholds are met). The FLM vehicle can serve multiple passengers on the last-mile leg but may pick up only one passenger on its way back. Hence, classifying a trip based on the number of passengers served can lead to some ambiguity. We, therefore, do not define trips as segments that start and end at the metro station but count them separately for the first- and last-mile legs. In other words, trips are passenger-serving segments between instances where the FLM vehicles are empty. The average number of trips serving one, two, and three passengers across all stations was found to be 777, 175, and 215, respectively.

%Path: Shared Folder Visualization/Tableau_AnyLogic_results.xlsx; Sheet name: "shareability_Stats"

FLM vehicles at stations such as Baiyyappanahalli, Majestic, and Indiranagar make a greater fraction of shared trips than the others. These stations have a high demand for boarding and alighting passengers, making it easier to pool rides. On the other hand, stations such as Peenya, Peenya Industry, and Goraguntepalya witness a lower fraction of shared trips because of their elongated Voronoi regions, which makes it difficult to pool rides. Sharing rides leads to lower lost demand, as demonstrated in Table \ref{tab:obj_value}. In addition, they also result in fewer vehicle kilometres travelled. Serving the passenger demand met in the joint FLM scenario using only single rides would have resulted in 406,434 vehicle km but allowing sharing reduces this to 295,036 vehicle km, a 27\% decrease, which decreases both congestion and emissions.

\subsection{Pricing Models}
The simulation framework proposed in this paper can also help analyse different pricing models, such as distance- and trip-based pricing schemes. In the following discussion, we compare revenue, assuming that the demand is captive and insensitive to the fare. The numbers used in this section are representative of auto-rickshaw operations in India. All calculations are estimated on a per-day basis.

\textbf{Distance-based pricing model}: To calculate the revenues, we first add up the fares paid by the set of passengers served at station $i$, $P(i)$. Passenger $p$ is assumed to be charged an amount $D_{ip}$ that depends on the direct distance between their origin/destination and the metro station. Metered auto-rickshaw fares in Bengaluru are of threshold type with a fixed base rate $\alpha$ up to 2 km and a distance-based price $\beta$ for subsequent kilometres travelled. Thus, the revenue from a passenger $p$ can be expressed as $R_{ip} = \alpha + \beta \max ( D_{ip}- 2, 0 )$, where $\alpha$ and $\beta$ were set to \rupee 30 and \rupee 15, respectively \citep{AutoFare}.

Let $D_{iv}$ be the distance travelled during an entire day by a vehicle $v$ assigned to station $i$. The operating cost is defined as $D_{iv} (\phi/\mu) + C_f$, where $\mu$ denotes vehicle mileage and $\phi$ is fuel price, set to 25 km per litre and \rupee 100 per litre, respectively. Operators are also assumed to incur a fixed cost $C_f$ set to \rupee 102 per vehicle per day \citep{Transport_Gov}, which includes depreciation and maintenance. For simplicity, $C_f$ is kept constant but may, in practice, depend on vehicle utilisation and age. Thus, the overall profit for FLM vehicles at station $i$ can be expressed as $P_i = \sum_{p \in P(i)} R_{ip} - \sum_{v = 1}^{x_i} \left( D_{iv} (\phi/\mu) + C_f \right)$.

%Path: Shared Folder Visualization/Tableau_AnyLogic_results.xlsx; Sheet names: "distance_based_vehicle_economic" and "consolidated_results"
\begin{figure}[H]
\centering
\includegraphics[scale=0.56]{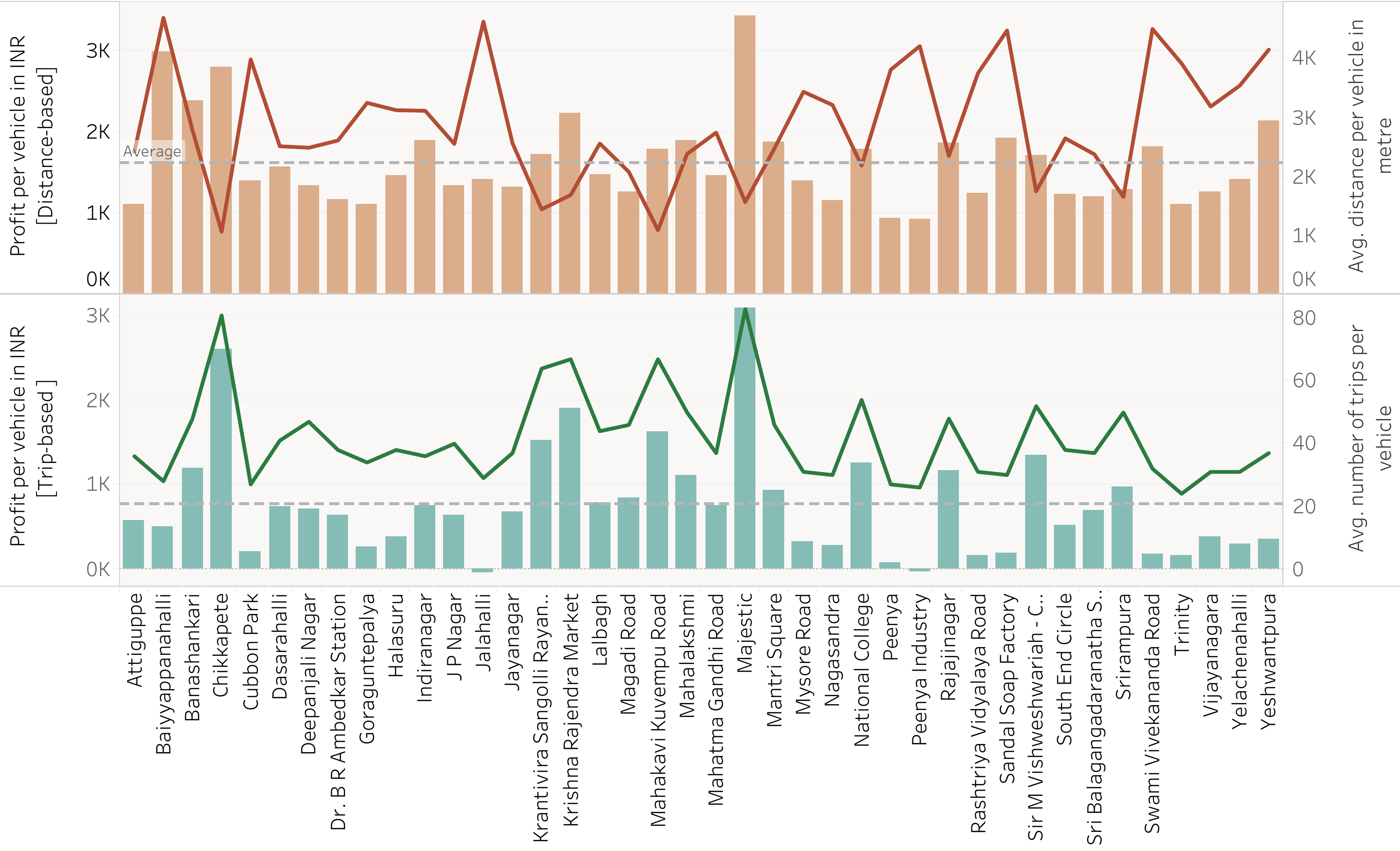}
\caption{Revenues under distance- and trip-based pricing.}
\label{fig:pricing_model}
\end{figure}

The top panel in Figure \ref{fig:pricing_model} shows the profit per vehicle at different metro stations under this pricing scheme. The average profit across all vehicles was found to be \rupee 1623. The line plot corresponding to the secondary axis shows the average distance travelled by the FLM vehicles. Compared to the daily earnings of \rupee 800--1200 reported in previous studies on auto-rickshaw drivers in Bengaluru city \citep{ramachanderfinancial}, we observe that serving metro passengers through a centralised service provider can yield higher profits. The estimated profits per vehicle at metro stations Baiyyappanahalli and Majestic are the highest as expected due to high passenger demand, vehicle utilisation, and degree of ride-sharing. Among all stations, Majestic has the smallest Voronoi region and, consequently, shorter average distances. However, since the pricing scheme has a base fare (independent of distance) for trips shorter than 2 km, the overall profits are considerably higher. On the other hand, Peenya Industry has the lowest profit per vehicle despite serving passengers with longer distances to the metro station. This result is possibly because Peenya Industry has the least passenger demand among all stations and a high share of single-passenger trips (see Figure\ref{fig:shareability_stats}).

\textbf{Trip-based pricing model}: We also estimated the profits from a trip-based pricing model where passengers are charged a flat fee of $R_{ip} =$ \rupee 30, irrespective of the distance they travel. The operating costs are assumed to be the same as before. As seen from the bottom panel of Figure \ref{fig:pricing_model}, the profit values are lower compared to the distance-based model. The average profit per vehicle was found to be \rupee 766, which is less than half of that in the distance-based case. At a fixed price of \rupee 44, the trip-based profits per vehicle become nearly equal to the estimates from the distance-based pricing model. Two metro stations -- Jalahalli and Peenya Industry -- have negative revenues due to a lower number of trips (see the line plot on the secondary axis). Setting a higher fixed fare for FLM vehicles at such stations could resolve this issue.

\subsection{Policy Implications}
Based on the above analysis, the following list summarises a few possible policy implications.
\begin{itemize}
\itemsep 0pt
    \item The marginal benefit of adding FLM vehicles to a station drops beyond a threshold. This threshold is different for each station and depends on the size and demand of the region it serves. Hence, service operators must determine an optimal number based on the trade-off between operating costs and net revenue.
    \item Since vehicle utilisation rates and profits vary significantly across stations, a policy of rotating vehicle assignments to stations may be required to provide equitable earnings to all drivers.
    \item There are temporal variations in lost demand, with peak hours witnessing three to four-times higher percentage of lost demand than off-peak hours. Considering the importance of public transit in reducing congestion, especially during peak periods, it is imperative to manage peak-period lost demand more efficiently. Ensuring maximum vehicle availability, augmenting the fleet size, and operating larger vehicles could reduce the demand lost during peak hours. 
\end{itemize}

\section{Conclusions}
\label{sec:conc}
The first- and last-mile parts of a trip are often inconvenient to commuters, which in turn results in low utilisation of public transport. One could develop a multimodal simulation for FLM connectivity to address this problem and test various improvement strategies before implementing such systems in the field. These simulations involve modelling the behaviour and interaction of agents such as passengers and FLM vehicles, arrival of metro trains, movement of vehicles on the road network, and pooling multiple passenger requests when shareability is allowed. 

We proposed a combination of agent- and event-based modelling and simulation approach to study a planning and an operational problem. At a planning level, we tried to improve the resource allocation of FLM services at different metro stations. We modelled the FLM vehicles and passengers as agents and described their behaviour using statecharts. Vehicles are assumed to be initially parked at metro stations, and routing is done using deterministic shortest paths. We developed an optimisation model that minimises the number of unserved trip requests (lost demand) by decomposing the problem by stations and formulating a knapsack-like problem with an approximate piecewise linear objective. We applied our model to the Bengaluru city metro transit network and carried out a day's simulation using real-world train schedules and passenger demand data.   

For the planning problem, our proposed method could allocate resources more efficiently, leading to a 26\% improvement in the objective compared to the best benchmark strategy. At an operational level, we tested different shareability scenarios and were able to quantify the benefits of sharing rides. For instance, an integrated FLM shareability scenario could reduce the lost demand by nearly 81\% and the vehicle km travelled by 27\%. Several other key performance indicators, such as percentage of lost demands on hourly and station-wise basis and fleet utilisation were explored, and suggestions for setting FLM fares were also made.  

While this research focuses on supply aspects and provides a framework for managing FLM fleet, many extensions that relax some assumptions can be conceived. First, the quality of FLM demand estimates and their origins/destinations can be improved with data on land use and employment characteristics; and trip types and their distribution during peak and off-peak periods. Second, modelling the effects of pricing and waiting times on multimodal demand is critical. In the context of pricing, many cities are exploring new forms of payment mechanisms, such as bundle pricing and revenue sharing, which can make multimodal journeys attractive. In addition, passengers could be given discounts for sharing rides; setting these fares at profitable levels is worthy of study. Capturing heterogeneity in waiting time depending on the alternate options available to passengers in different geographical areas of a city can also improve model fidelity. Third, while zonal restrictions for FLM vehicles alleviate the need for rebalancing, a certain degree of optimality is lost when there are opportunities to share trips by passengers close to the boundaries. Understanding these trade-offs without making the simulation unmanageable is another promising direction for future research. Fourth, incorporating time-of-day changes in traffic and congestion effects can improve the estimates of ride-sharing benefits. Since our results indicate significant differences in vehicle utilisation across the day, one could also explore dynamic supply allocation strategies to optimise expenditure. Finally, anticipating the electrification of shared vehicle fleets, adjusting FLM operations to accommodate recharging or battery swapping can broaden the applicability of this decision support simulator. 

\section*{Acknowledgements}
This study is supported by IMPacting Research, INnovation and Technology (IMPRINT), Department of Science and Technology (DST), India (Project no. IMP/2018/001850). The authors thank Bangalore Metro Rail Corporation Limited (BMRCL) for sharing their data. Assistance from Ms. Deepa L on collating census data is also appreciated.  

This preprint has not undergone peer review or any post-submission improvements or corrections. The Version of Record of this article is published in \textit{Transportation}, and is available online at \url{https://doi.org/10.1007/s11116-022-10363-z}.

\bibliography{references}

\newpage
\appendix
\section{AnyLogic Specifics}
\label{appendix:specs}
For the purpose of reproducibility, this subsection provides a few implementation details that are specific to AnyLogic. These details are not critical for simulation platforms built using other tools, and hence the reader can skip this material without any loss of continuity. 

\begin{figure}[H]
\centering
\includegraphics[scale=0.65]{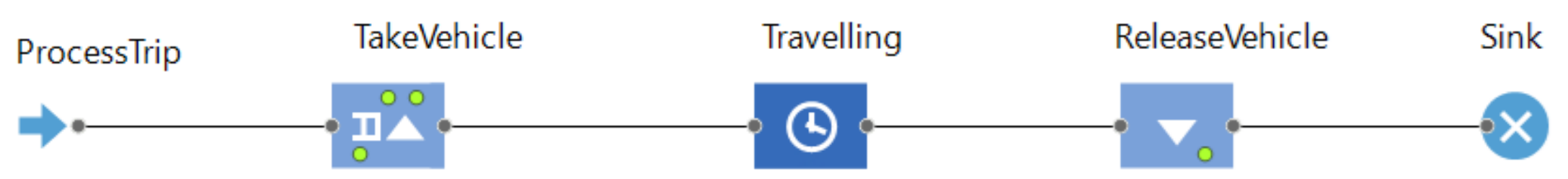}
\caption{Process modelling library blocks}
\label{fig:process_modelling_lib_blocks}
\end{figure}

The discrete-event model that triggers the occurrence of certain time-ordered processes \citep{borshchev2004system} is executed in AnyLogic using its \textit{process modelling library} blocks. Individual events can also have a timestamp associated with them and can be scheduled in advance. To create vehicle agents for every metro station, we use the \textit{resource pool} functionality of AnyLogic's process modelling library. The number of resource units in each pool can be specified along with other properties such as vehicle speed and initial location. An advantage of using resource pools is that it captures utilisation statistics, such as the time for which FLM vehicles are idle or busy during the simulation. These can be computed using a set of process modelling library blocks as shown in Figure \ref{fig:process_modelling_lib_blocks}. When an instance \texttt{trip} (agent type \texttt{LastMileTrip} or \texttt{FirstMileTrip}) is generated, the function \texttt{ProcessTrip.take(trip)} is called. \texttt{ProcessTrip} is an \texttt{Enter} block at which the FLM vehicle agents are inserted into the process modelling library blocks. \texttt{TakeVehicle} is a \texttt{Seize} block that captures units of the FLM vehicle resource pool. The control then goes to the \texttt{Travelling} block, which is a \texttt{Delay} block, until the \texttt{stopDelay()} function is called after an FLM vehicle serves the passenger trip request and reaches the station. After this step, the control flow goes to the \texttt{ReleaseVehicle} block that releases the seized FLM vehicle unit. The agents are then taken out of this process flow by the \texttt{Sink} block. 
    
For simulating first-mile passenger arrivals, the \textit{dynamic event} functionality of AnyLogic was used (since it allows the creation of multiple concurrent instances, each of which is independent) and initialised with the specified parameters (such as the name of the metro station where passengers arrive). For the shareability scenarios, communication of inputs and outputs between the CVRP codes and AnyLogic is established through additional custom code using \textit{PypeLine} (a Python connector library for AnyLogic).

\end{document}